\documentclass{emulateapj}
\usepackage{graphics}
\usepackage{epsfig}
\usepackage{subfigure}

\newcommand{\ltsimeq}{\la}
\newcommand{\gtsimeq}{\ga}
\newcommand{\msun}{M$_{\odot}$}
\newcommand{\zsun}{Z$_{\odot}$}

\newcommand{\HII}{H~{\sc ii}}

\shortauthors{McQuinn, Skillman et al.}
\shorttitle{Observational Constraints on Red and Blue Helium Burning Sequences}

\begin{document}
\submitted{July 2011}
\title{Observational Constraints on Red and Blue Helium Burning Sequences\footnote{Based on observations made with the NASA/ESA Hubble Space Telescope, obtained from the Data Archive at the Space Telescope Science Institute, which is operated by the Association of Universities for Research in Astronomy, Inc., under NASA contract NAS 5-26555.}}
\author{Kristen B.~W.~McQuinn\altaffilmark{1} 
Evan D.~Skillman\altaffilmark{1}
Julianne J.~Dalcanton\altaffilmark{2},
Andrew E.~Dolphin\altaffilmark{3},
Jon Holtzman\altaffilmark{4},
Daniel R.~Weisz\altaffilmark{2}, \&
Benjamin F.~Williams\altaffilmark{2}
}

\altaffiltext{1}{Department of Astronomy, School of Physics and
Astronomy, 116 Church Street, S.E., University of Minnesota,
Minneapolis, MN 55455,\ {\it kmcquinn@astro.umn.edu}} 
\altaffiltext{2}{Department of Astronomy, Box 351580, University 
of Washington, Seattle, WA 98195}
\altaffiltext{3}{Raytheon Company, 1151 E. Hermans Road, Tucson, AZ 85706}
\altaffiltext{4}{Department of Astronomy, New Mexico State University, Box 30001-Department 4500, 1320 Frenger Street, Las Cruces, NM 88003}

\begin{abstract}

We derive the optical luminosity, colors, and ratios of the blue and red helium burning (HeB) stellar populations from archival Hubble Space Telescope observations of nineteen starburst dwarf galaxies and compare them with theoretical isochrones from Padova stellar evolution models across metallicities from $Z=0.001$ to 0.009. We find that the observational data and the theoretical isochrones for both blue and red HeB populations overlap in optical luminosities and colors and the observed and predicted blue to red HeB ratios agree for stars older than 50 Myr over the time bins studied. These findings confirm the usefulness of applying isochrones to interpret observations of HeB populations. However, there are significant differences, especially for the red HeB population. Specifically we find$\colon$ (1) offsets in color between the observations and theoretical isochrones of order 0.15 mag (0.5 mag) for the blue (red) HeB populations brighter than M$_V\sim -4$ mag, which cannot be solely due to differential extinction; (2) blue HeB stars fainter than M$_V \sim -3$ mag are bluer than predicted; (3) the slope of the red HeB sequence is shallower than predicted by a factor of $\sim3$; and (4) the models overpredict the ratio of the most luminous blue to red HeB stars corresponding to ages $\ltsimeq50$ Myr. Additionally, we find that for the more metal-rich galaxies in our sample ($Z\gtsimeq 0.5$~\zsun) the red HeB stars overlap with the red giant branch stars in the color magnitude diagrams, thus reducing their usefulness as indicators of star formation for ages $\gtsimeq100$ Myr.
\end{abstract} 

\keywords{galaxies: dwarf --- galaxies: evolution --- galaxies: individual (Antlia dwarf, ESO~154-023, UGC~4483, UGC~6456, UGC~9128, NGC~625, NGC~784, NGC~1569, NGC~2366, NGC~4068, NGC~4163, NGC~4214, NGC~4449, NGC~5253, NGC~6789, NGC~6822, IC~4662, DDO~165, Holmberg~II) --- galaxies: starburst}

\section{HeB Stars in Isochrones and Observational CMDs }\label{intro}

Theoretical stellar evolutionary isochrones predict the paths that stars will traverse in a color magnitude diagram (CMD) as the stars age and evolve. Isochrone libraries span a large range of stellar masses and metallicities \citep[e.g.,][and references therein]{VandenBerg2000, Bertelli1994, Lejeune2001, Yi2001, Pietrinferni2004, Dotter2008, Marigo2008}. They are widely used to aid in the interpretation of CMDs constructed from stellar populations observed in clusters and galaxies including measuring the star formation histories (SFHs) of galaxies \citep[e.g.,][]{Tolstoy1996, Aparicio1996, Dolphin2002, Harris2001}. Isochrones have also aided both in understanding of the integrated properties of stellar populations \citep[e.g.,][and references therein]{Bruzual2003, Leitherer2010} and in quantifying the chemical and spectral evolution of galaxies. Current theoretical isochrones use radiative opacities \citep{Bertelli1994} and modified combinations of convective core and convective envelope overshooting \citep{Alongi1993,Bressan1993} calibrated to the different wavelengths and filters commonly used in observations \citep{Girardi2004}. 

In this paper we focus on luminous, core helium burning (HeB) stars. These stars have initial masses between $\simeq2 - 15$ \msun, and evolve off the main sequence (MS) between ages of $\simeq5 - 1000$ Myr into HeB stars before evolving further into thermally pulsating asymptotic giant branch (TP-AGB) stars or exploding as supernovae. Specifically, current models show stars in this initial mass range will leave the MS burning helium in their cores \citep[e.g.,][and references therein]{Gallart2005}, becoming more luminous and redder in color, and will populate the red HeB (RHeB) branch in a CMD. The RHeB stars will traverse the CMD  toward bluer colors, making the ``blue loop'', and populate the blue HeB (BHeB) branch. A comparison of the predicted characteristics of HeB stars with observations can provide important constraints for the stellar evolution models, particularly with high-fidelity photometric measurements now achievable from space-based telescope observations. 

Based on the current Padova models \citep[for a comprehensive review, see][]{Chiosi1992}, both the RHeB and BHeB phases of evolution occupy a space in an optical CMD which is uniquely dependent on the age of the star. While other phases of evolution have either an age degeneracy or an age-metallicity degeneracy in their CMD location, the one-to-one age-luminosity relation of HeB stars affords an opportunity to not only separate stars in this stage of evolution from other populations, but to also age date the HeB population in an observed CMD directly from isochrones \citep[e.g.,][]{Dohm-Palmer1997}. HeB stars spend more time at or near the extreme red and blue colors than at intermediate colors as HeB stars evolve rapidly from the RHeB to the BHeB phase. The luminosities and colors around these extrema define the RHeB and BHeB sequences observationally and are readily identifiable in an observed CMD. The equivalent red and blue extrema are also easily identified in theoretical isochrones. Thus, the HeB populations provide an excellent opportunity to test the predictions of theoretical isochrones on an individual stage of stellar evolution. In addition, the ratio of BHeB stars to RHeB stars, or B/R ratio, is now measurable for resolved stellar populations. The B/R ratio is dependent on convection, mass-loss, and stellar rotation, and thus also provides important diagnostics for stellar evolutionary models \citep[e.g.,][and references therein]{Dohm-Palmer2002}.

To illustrate the importance of HeB stars for constraining stellar evolution, one can consider the historical development of stellar models with an initial mass between $\simeq2 - 15$ \msun. For these stars, it was originally assumed that the sole opacity source was electron scattering, which was later revised to include bound-free absorption \citep{Hayashi1962a, Stothers1966}. Other studies assumed the unstable part of the stellar envelope of massive stars was fully convective \citep{Iben1966} instead of semi-convective. Both the opacity assumption and convection assumption predicted that massive stars (M $>15$ \msun) would evolve off the MS directly into blue supergiants or BHeB stars. Red supergiants were thought to have a different origin and were explained as a later phase of evolution of stars with carbon-burning cores \citep{Hayashi1962b}. 

The subsequent addition of bound-free absorption and semi-convective envelopes in the stellar evolution models \citep{Stothers1968} changed our understanding of the evolutionary paths of the massive stars; i.e., they correctly predicted that \textit{both} blue and red supergiants are core HeB stars, and that MS stars evolve first into RHeB stars and then into BHeB stars. With the blue and red supergiants both identified as core HeB stars, the RHeB stars were predicted to traverse the CMD as they evolve into BHeB stars when the mass fraction of hydrogen shell burning increased and the efficacy of cooling by bound-free absorption was reduced. Revisions of the triple$-\alpha$ nuclear reaction rate induced the formation of a blue loop for all but the most massive stars \citep{Austin1971}.  Further, the blue loop was lengthened due to changes in the cross-section of the $^{12}$C($\alpha,\gamma$)$^{16}$O nuclear reaction \citep[see \S6.3][and references therein]{Gallart2005}. 

The predicted ratio of the blue to red (B/R) HeB stars (i.e., predictions of the evolutionary time spent in each core HeB phase) was found to depend on the prescription for convection, with the Ledoux criterion \citep[][and references therein]{Ledoux1947, Sakashita1959, Stothers1975} better reproducing the general trends in the distribution of blue and red supergiants than the Schwarzschild criterion \citep[][and references therein]{Stothers1976}. The inclusion of downward convective envelope overshooting more consistently predicted the occurrence of blue loop stars \citep{Stothers1991} favoring these models over those including convective core overshooting, which under-predicted the blue colors of the BHeB stars \citep{Bertelli1985, Stothers1992} or over-predicted the luminosity of the BHeB stars \citep[][and references therein]{Alongi1993}. 

In spite of these successes, discrepancies in the models of the HeB stars remained. Further study of RHeB stars revealed that the stellar evolution models that correctly predicted the number of red supergiants at low metallicities over-predicted the corresponding number of RHeB stars in solar metallicity systems and vice-versa \citep[e.g.,][and references therein]{Langer1995}. In response, stellar rotation was added to the models, increasing internal mixing and causing a larger He-core. The smaller associated intermediate convective zones permitted the stellar radius of rotating stars to inflate during the He-burning phase, favoring the evolution of stars towards red supergiants \citep{Maeder2001}. This increased the number of RHeB stars, thus producing B/R ratios more consistent with low metallicity observations. However, accurately measuring the B/R ratio from ground based observations of this era was difficult as the lower photometric accuracy blended the MS and BHeB stars into the ``blue plume'' of an optical CMD.

Using observations from the Hubble Space Telescope (HST) archive, we expand on the work by \citet{Dohm-Palmer2002} for Sextans~A and measure the luminosity, color, and ratios of the optically resolved BHeB and RHeB stellar sequences in the CMDs of nineteen starburst dwarf galaxies. Starburst galaxies were selected as they are known to contain significant populations of HeB due to their recent and/or ongoing star formation. The reduction in uncertainty due to the large number of HeB stars is somewhat offset by the associated increase in uncertainty due to the higher crowding and differential extinction typical in regions of high star formation \citep{Boquien2009}. The galaxy sample spans a range of chemical composition from $Z = 0.0007$ to 0.0093 allowing us to test isochrones of metallicities up to $\simeq0.5$ \zsun. We compare the measured sequences and the B/R ratios to those derived from Padova stellar evolution isochrones. The Padova isochrones were selected for comparison because they are currently the only isochrones with sufficient metallicity and high mass star coverage, and, conveniently, the isochrones are calibrated to the HST filters used for the observations.

In \S\ref{theoretical}, we present the theoretical isochrones and describe how the predicted BHeB and RHeB sequences were identified. In \S\ref{empirical}, we describe the galaxy sample and how the sequences were identified from the observational data. In \S\ref{compare} we compare the theoretical and empirical results. We present a discussion of the impact of these results in \S\ref{discussion} and our conclusions in \S\ref{conc}.

\section{Measuring the BHeB and RHeB Sequences of Padova Isochrones }\label{theoretical}

The predicted luminosity, color, and ratio of HeB stars of different ages (i.e., masses) can be discerned from theoretical isochrones. We use the theoretical isochrones of \citet{Girardi2000} for stars with masses $\leq 7$ \msun\ with improved treatment of the asymptotic giant branch (AGB) stars added by \citet{Marigo2008}. At higher masses (M$>$7 \msun), we use the isochrones of \citet{Bertelli1994}. Collectively, this set of isochrones is known as the Padova stellar evolution isochrones. The Padova isochrones (hereafter ``isochrones'') were selected for comparison with observations because they met three criteria. First, the isochrones are available for the range in metallicities of the galaxy sample. Second, the Padova stellar evolution models include high mass stars that populate the blue and red HeB sequences. Third, the isochrones provide predicted magnitudes calibrated to the HST filters and are thus directly comparable to the observational data used in this study. The basic outputs of the stellar evolution models are surface luminosity and effective temperatures ($T_{eff}$) of stars. To allow comparison with observed photometric data, these quantities are converted to magnitudes and color for various bandpasses using bolometric corrections and $T_{eff}$-color relations from stellar atmosphere models, namely the ATLAS9 stellar spectra library \citep{Castelli2003}. For a complete description of the stellar atmosphere libraries used, see \citet{Girardi2008} and references therein. We chose not to compare the data to the BaSTI models \citep{Pietrinferni2004} as the isochrones do not cover include the higher mass HeB stars. Similarly, we chose not to use the updated Geneva models \citep[e.g.,][]{Meynet1997} as the isochrones cover only metallicities of $Z=0.00001, 0.004$, or solar abundance. 

Isochrones of seven different chemical compositions were selected with $Z$ ranging from $0.001-0.019$. The six lowest metallicity isochrones approximately match the metallicities of our galaxy sample as listed in Table~\ref{tab:galaxies}. The highest metallicity is equivalent to solar abundance, and is included for comparison purposes. We generate these seven sets of isochrones for three different permutations of HST instruments and filters, listed in Table~\ref{tab:instr_filter}. While our data were not obtained with the bluer HST filter F475W, we include predictions for the F475W and F814W combination for comparison purposes and future reference. We chose not to use circumstellar dust models \citep{Bressan1998, Groenewegen2006} which incorporate mass loss to asymptotic giant branch (AGB) and post-AGB stars as the dust models impact the signatures of AGB stars and not HeB stars.

\begin{figure}
\epsscale{1}
\plotone{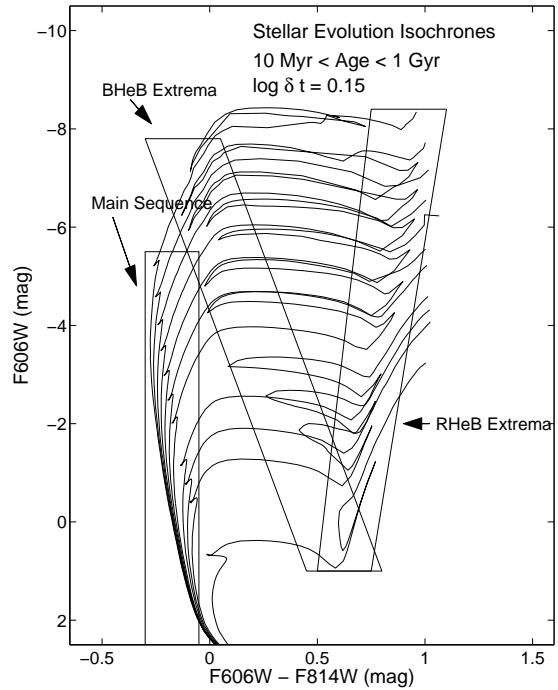}
\caption{0.003 isochrones for stars with ages between log t$ = 7.0$ to 9.0 at increments of log $\delta$ t $=0.15$ for the F606W and F14W filters. Three evolutionary sequences are identified in the figure: the main sequence of stars, the red extrema of the HeB stars occurring after the stars leave the main sequence, and finally the blue extrema of the HeB stars reached after the RHeB stars traverse back toward bluer colors in the CMD. The isochrones are cut-off after the BHeB stage of evolution to avoid confusion in the plot.}
\label{fig:cmd_iso}
\end{figure}

In Figure~\ref{fig:cmd_iso}, we present an example of the theoretical isochrones used in our analysis. The isochrones for a single metallicity of 0.003 are plotted for one set of V- and I-band filters, for stellar ages from log~$t = 7.0 - 9.0$ at increments of log $\delta$t$ = 0.15$. Each line traces the CMD location of stars of the same age but with different masses, thus showing the position of main sequence (MS) stars, red and blue HeB stars, and their evolution off the CMD at the end of the life cycle. We do not show isochrones for ages greater than 1 Gyr, i.e., for lower mass stars ($\ltsimeq 2$ \msun) that would evolve onto the red giant branch (RGB) without becoming HeB stars. Note that the BHeB stars have values of $M_V \sim 2$ mag brighter than MS stars of equal mass.

\begin{figure}
\epsscale{1}
\plotone{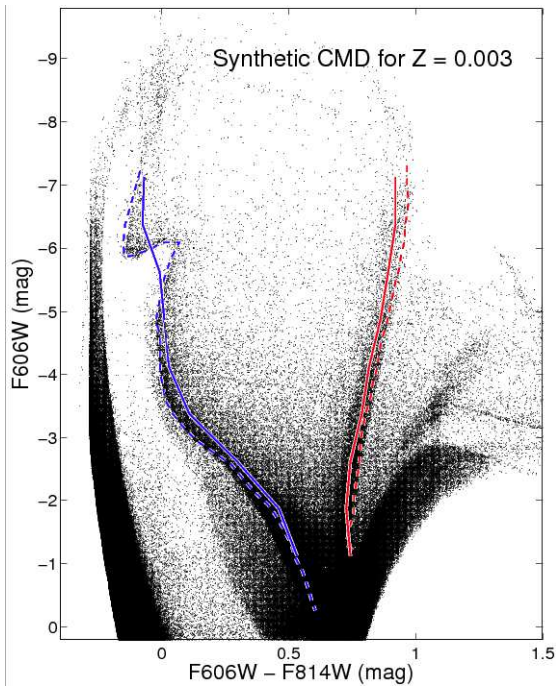}
\caption{A synthetic CMD generated for a metallicity of $Z=0.003$. The model assumes increasing SFRs at recent times to fully populated the CMD with higher luminosity, higher mass stars. The two dashed lines represent the blue and red extrema of the HeB tracks in the isochrones. The blue and red HeB stars spend most of their time approaching the extrema, thus the HeB stars in the synthetic CMD lie near the isochrone predicted sequences but with some distribution in color. We determine the luminosity and color of the HeB sequences from the synthetic CMDs, given that the observed HeB sequences will `pile-up' in a similar fashion as they approach the extreme colors of the blue loop. The adopted sequences are shown as solid lines.}
\label{fig:cmd_synth}
\end{figure}

HeB stars will be found at colors and magnitudes which approximately correspond to the red and blue extrema of the isochrones, corresponding to the beginning of the core HeB stars and the maximum temperature of the BHeB phase. As HeB stars spend more time near these extrema than at intermediate colors and magnitudes, the HeB stars create two loci of points or ``sequences'' in the CMD (see \S\ref{emp_data} Figure~\ref{fig:cmd_data}, for example). However, when weighted by the duration the stars spend in the HeB phases, these sequences do not necessarily correspond exactly to the extrema. It is the luminosity and color of the actual blue and red HeB sequences (i.e., not the extrema of the isochrones) that correspond to the sequences in an observed CMD and thus of interest here. Thus, to determine the exact location of the sequences, we generated synthetic CMDs using seven different metallicity isochrones, assuming a Salpeter initial mass function \citep{Salpeter1955} for two different V and I filter combinations. An example synthetic CMD is shown in Figure~\ref{fig:cmd_synth} for the isochrone $Z =0.003$ and the F606W and F814W filter combination.

\begin{figure}
\epsscale{1}
\plotone{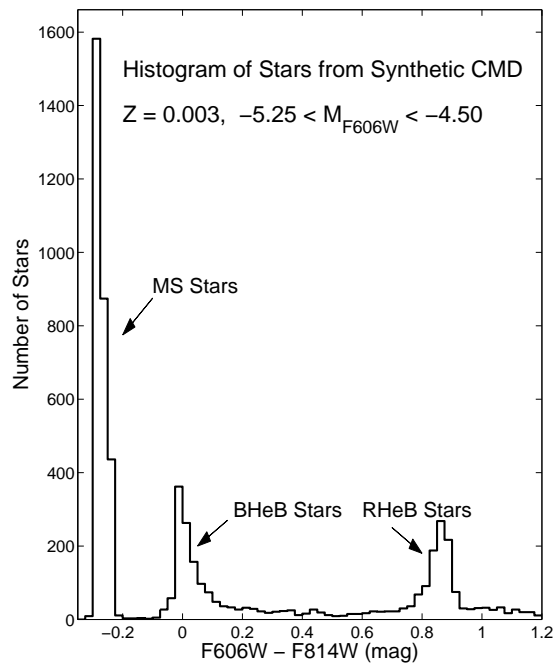}
\caption{An example histogram of stars from the synthetic CMD in Figure~\ref{fig:cmd_synth} between the luminosity range $=5.25<$M$_V <-4.50$. The first peak in the histogram is populated by MS stars, the second peak by BHeB stars, and the final peak by RHeB stars. The color separation between the difference sequences is evident in the figure allowing the luminosity and color of the HeB sequences to be unambiguously identified. We chose the peak of the blue and red HeB histogram peaks to define the sequences.}
\label{fig:histo_synth}
\end{figure}

To determine the locations of these sequences in the CMDs, we binned the stars in the synthetic CMD in color over nine one-magnitude wide luminosity intervals. We present an example histogram of a one-magnitude slice in Figure~\ref{fig:histo_synth}. The three peaks in each histogram correspond to the MS stars, the BHeB stars, and the RHeB stars. The theoretical BHeB sequence was determined by the colors of BHeB distribution peaks at each of the nine one-magnitude intervals in luminosity. The RHeB sequence was determined using the same method, but over a smaller magnitude range as RHeB stars do not reach the same high luminosities as the BHeB stars at optical wavelengths. We plot these sequences as solid lines on the synthetic CMD in Figure~\ref{fig:cmd_synth} and compare them to the location of the blue and red extrema shown as the dashed lines. As expected, the histogram-determined sequences lay on the centers of the blue and red HeB stellar distributions, slightly offset from the extrema.

\begin{figure}
\epsscale{1}
\plotone{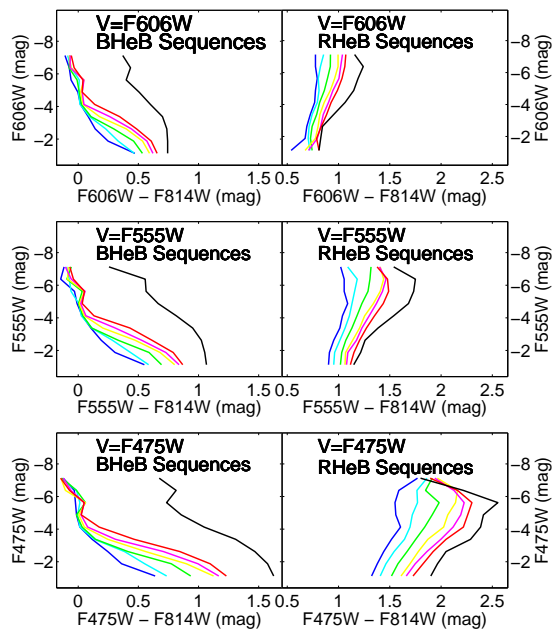}
\caption{Theoretical BHeB and RHeB sequences for three filter combinations. The colors represent different isochrone metallicities matching the colors and metallicity range of the observed galaxy sample shown in Figure~\ref{fig:emp_iso}. Color key: blue is $Z=0.001$, cyan is $Z=0.002$, green is $Z=0.003$, yellow is $Z=0.005$, magenta is $Z=0.006$, red is $Z=0.009$, black is $Z=0.019$. The black sequence represents solar abundances and is added for comparison purposes. Note that the small shifts in the V filter between F606W, F555W, and F475W, causes a measurable shift in the colors of the sequences. }
\label{fig:theory_iso}
\end{figure}

In Figure~\ref{fig:theory_iso}, we present the theoretical predictions for the blue and red HeB sequences for seven metallicities and different optical filter combinations. The top panels show the blue (left) and red (right) HeB track for the F606W and F814W filters for the HST  Advanced Camera for Surveys \citep[ACS;][]{Ford1998}. The colored lines represent different metallicities. The lowest metallicity ($Z =0.001$; blue line) has the overall bluest colors for both the blue and red HeB sequences. The highest metallicity ($Z =0.019$ or solar abundance; black line) has the reddest colors for the blue and red HeB sequences. The middle panels of Figure~\ref{fig:theory_iso} show the HeB tracks for F555W and F814W filters for the HST Wide Field Planetary Camera 2 \citep[WFPC2;][]{Holtzman1995} instrument. The HeB sequence generated from isochrones for the ACS F555W and F814W filters were equivalent to those generated for the equivalent WFPC2 filters. We therefore use the isochrone set from the WFPC2 filters for comparison with observations from both instruments and refer to the isochrone set by filter, not by instrument. The bottom panels of Figure~\ref{fig:theory_iso} show the HeB tracks generated for F475W and F814W filters for the HST ACS instrument. The isochrones using this longer color baseline cover a larger range in color in the CMDs.

\section{Measuring the BHeB and RHeB Sequences in the Observed CMDs }\label{empirical}

\subsection{The Galaxy Sample and Their Optically Resolved Stellar Populations }\label{emp_data}

The galaxy sample was comprised of nineteen, nearby starburst dwarf galaxies, previously studied by \citet[][and references therein]{McQuinn2010b}. All of the galaxies have high quality HST multi-color optical imaging taken with either ACS or WFPC2, previously reduced and photometered as described in \citet{McQuinn2010a, McQuinn2010b}. This sample forms an ideal set for identifying HeB sequences as the recent rates of star formation are sufficiently high such that the blue and red HeB branches of the observed CMD are well-populated. While there is some associated increase in uncertainty in a starburst galaxy sample due to the higher crowding typical in regions of high star formation, most of our sample is comprised of low metallicity dwarf galaxies that have relatively small amounts of differential extinction. Thus, our choice in galaxies reduces, although does not eliminate, the effects of differential extinction and the potential for blending of MS and BHeB stars in a CMD, in spite of our sample's high recent star formation rates (SFRs) and high gas content.

Individual metallicities for the sample are listed in Table~\ref{tab:galaxies}. The metallicities are derived either from \HII\ region abundances taken from the literature or from estimates derived from the luminosity-metallicity relation \citep{Zaritsky1994, Tremonti2004, Lee2006}, as noted. The present-day metallicities are likely to match those of the HeB populations, given that no significant chemical evolution is expected or observed over the last $\sim250$ Myr. Studies of the oxygen abundances in \HII\ regions show the gas is well mixed in dwarf galaxies on short timescales \citep{Skillman1989, Kobulnicky1997b, Lee2006}, thus any variations in metallicity in the young stellar populations within an individual galaxy should be negligible.

\subsection{Correcting the Observations for Reddening }\label{de-redden}

The Galactic foreground and internal extinction in the V filter was estimated through fitting the observed CMD to synthetic CMDs as described in \citet{McQuinn2009, McQuinn2010a}. Briefly, we reconstructed the SFHs by finding the best fit synthetic CMD to an observed CMD allowing the distance and the average uniform foreground extinction to be included as parameters. A differential extinction parameter was added in cases where the observed CMD showed signs of internal extinction, typically manifested as broadening of the MS beyond what is expected from photometric uncertainties. We list our total estimated extinction, including both foreground and internal extinction values, in Table~\ref{tab:galaxies} and compare them with the foreground extinction estimates from the dust maps of \citet{Schlegel1998}. In sixteen cases, our estimates are within 0.1 mag of those from \citet{Schlegel1998}. In three cases, our estimates of the total extinction are greater than those of \citet{Schlegel1998} by $0.19-0.32$ mag, indicating the presence of internal extinction. Including estimates of internal extinction optimizes the chance for the models to fit the data, thus any significant discrepancies between the models and data in these three systems is more likely attributable to model parameters. 

The extinction for the F814W filter was determined by extrapolating our estimated $A_V$ values using the York Extinction Solver tool \citep{McCall2004} which applied the reddening law of \citet{Cardelli1989} with R$=3.1$. These $A_{F814W}$ values are also listed in Table~\ref{tab:galaxies}. We adopt our estimated constant values for de-reddening and account for the extinction by subtracting our estimated $A_V$ values from the stellar photometry and our estimated $A_V - A_{F814W}$ values from the measured $V-I$ stellar colors. The extinction reported in Table~\ref{tab:galaxies} does not account for the potential presence of differential extinction that may be located around the youngest stars; we consider the effects on the observed blue and red HeB sequences of such potentially non-uniform extinction in \S\ref{compare_bheb} and \S\ref{compare_rheb}.

\subsection{Isolating the HeB Populations }\label{bheb}

The BHeB stars populate a region of the optical CMD approximately parallel in magnitude to the MS stars. As long as the photometry are of high quality, the photometric errors at the brighter magnitudes of the HeB stars are smaller than the inherent separation in color between the two populations and the two distinct populations are easily separated in color-magnitude space. There is some contamination of the BHeB sequence as stars leaving the MS and evolving into RHeB stars pass through this loci of points. However, these stars spend very little time making the transition and are conservatively estimated to produce a contamination of less than $\sim10$\% \citep{Dohm-Palmer2002}. Likewise, the RHeB stars form a sequence that lies blue-ward of the RGB.

\begin{figure}
\begin{center}
\plottwo{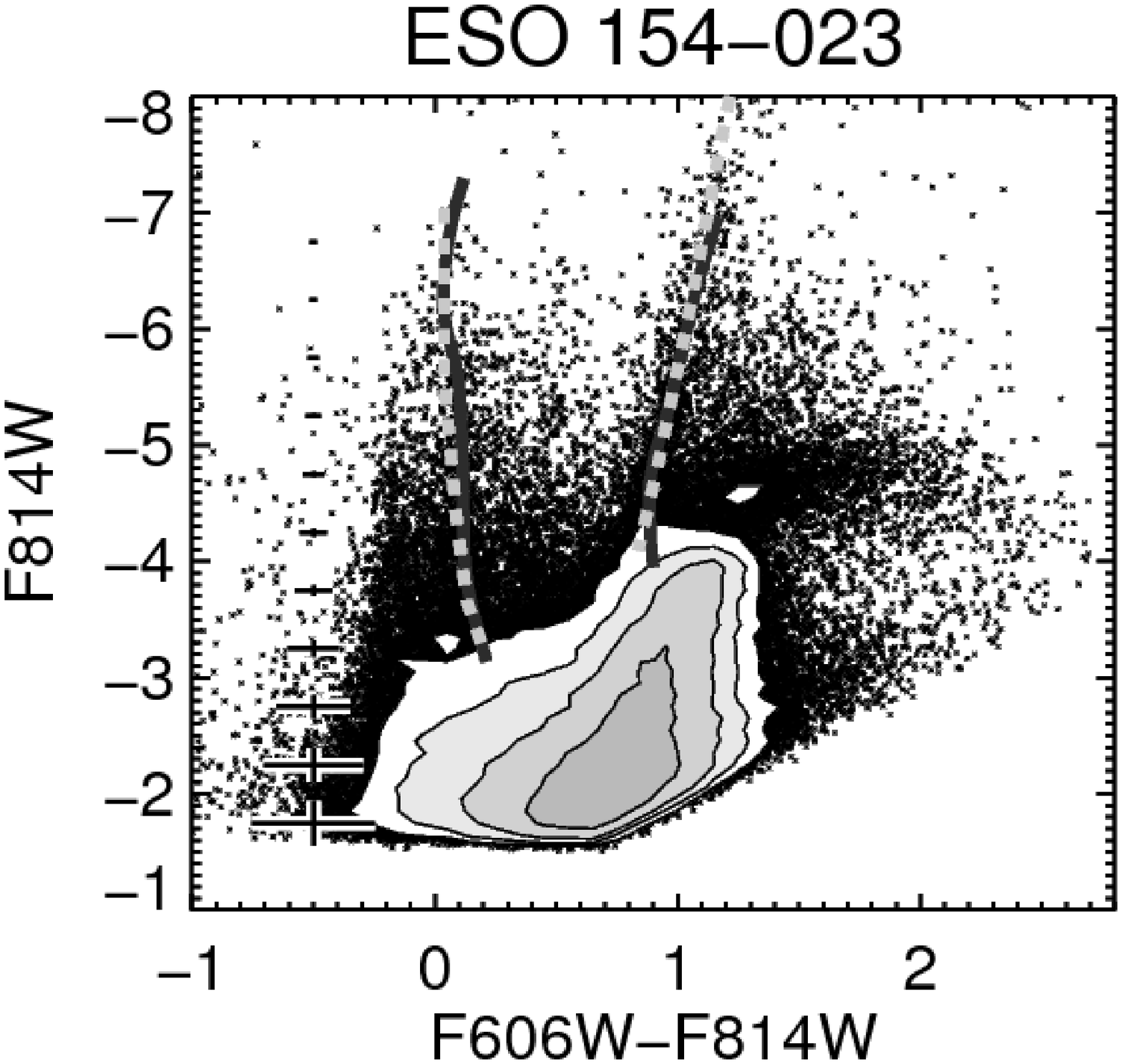}{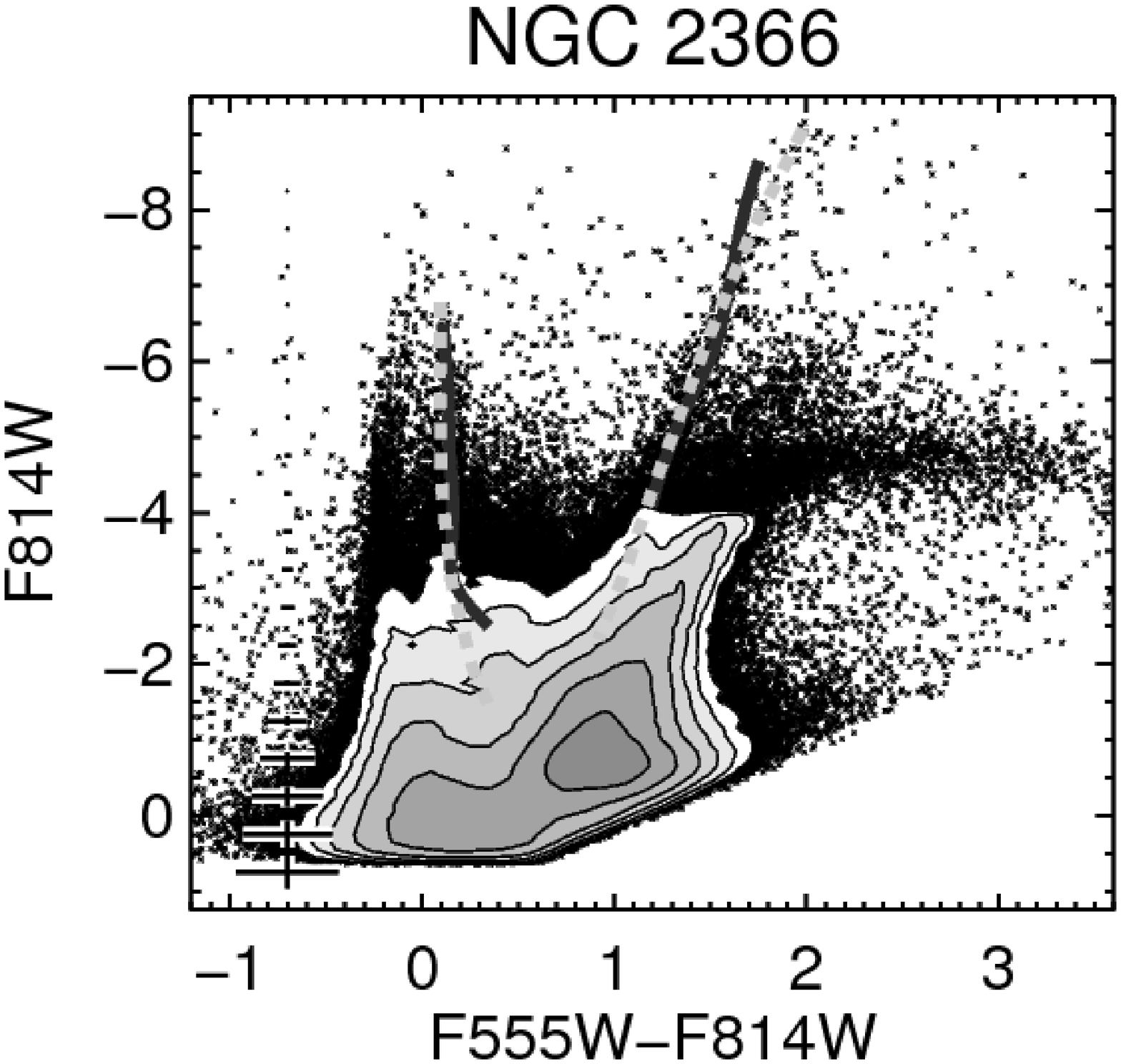}
\end{center}
\caption{Optical CMDs of a galaxy reaching a photometric depth representative of the sample (top panel; ESO154$-$023), and of a galaxy reaching one of the deepest photometric depths of the sample (bottom panel; NGC~2366). The dashed lighter lines denote the blue and red HeB sequences selecting by eye on the CMD. The solid darker lines are the same HeB sequences selected using histogram plots of stars in one-magnitude intervals as shown in Figure~\ref{fig:histo_data}. The blue and red HeB sequences selected with the two methods are in good agreement with each other.}
\label{fig:cmd_data}
\end{figure}

We take two approaches to selecting the loci of both the BHeB and the RHeB stars in the observational CMDs to ensure the most robust selection of HeB stars with the smallest contamination from MS or RGB stars. Here, we describe the two methods for selecting the BHeB stars in the CMDs; these same methods are applied in selecting the RHeB stars. First, the center of the BHeB stellar distribution is marked in the CMD, shown as a dashed line in the CMDs of Figure~\ref{fig:cmd_data} for two galaxies in our sample. The over-density of stars along these lines are apparent by eye. The spread in color of the BHeB populations is due to photometric uncertainties, differential extinction, and the longer periods of time BHeB spend near the blue extrema in their evolution. Note that systems with significant amounts of differential extinction blend the MS stars with BHeB stars. In contrast, the galaxies in our sample all show this over-density of BHeB clearly separable from the MS in their CMDs, making selection of the BHeB stars viable. 

\begin{figure}
\plotone{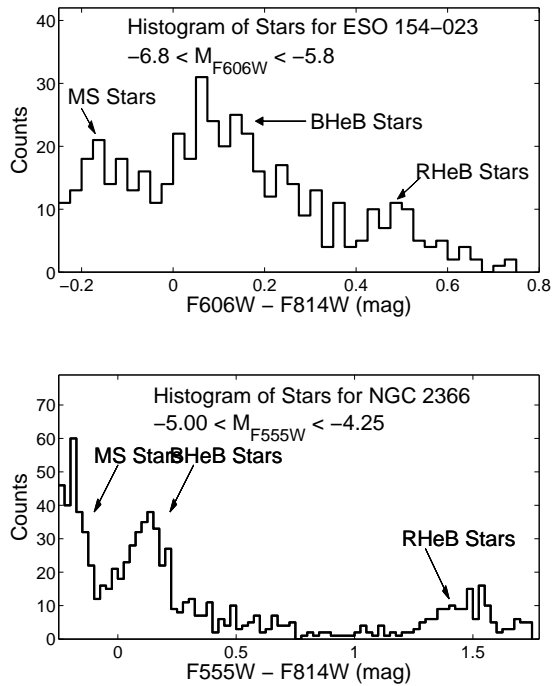}
\caption{Example histograms of the stars in ESO154$-$023 (top) and NGC~2366 (bottom) in a magnitude slice of the CMDs shown in Figure~\ref{fig:cmd_data} to be compared with theoretical histograms in Figure~\ref{fig:histo_synth}. The three peaks in the histograms correspond to the MS, BHeB, and RHeB populations. The color separation between the difference sequences is evident in the figure allowing the luminosity and color of the HeB sequences to be unambiguously identified. We chose the peak of the blue and red HeB histogram peaks to define the sequences.}
\label{fig:histo_data}
\end{figure}

Our second approach to isolating BHeB and RHeB stars follows the same methodology used to isolate the BHeB in the synthetic CMDs (see \S\ref{theoretical}). We generate histograms of color for all stars in the MS, BHeB, and RHeB regions in magnitude intervals varying from 0.5 mag to 1.0 mag, depending on the photometric uncertainties and SFH of each galaxy. Six magnitude intervals were used for the data instead of the nine used for the synthetic CMDs, as the location of the observed HeB sequences are measurable over a smaller range in luminosity than the synthetic HeB sequences. We present two example histograms in Figure~\ref{fig:histo_data} for data of shallower photometric depth (top panel; ESO~154$-$023) and for data of deeper photometric depth (bottom panel; NGC~2366). The histograms show the number of stars at each $V-I$ color in a sample magnitude interval. In each panel, there are three peaks corresponding to the MS, BHeB, and RHeB stars. 

We compared the colors and luminosities of the BHeB and the RHeB sequences that were defined by our two different methods. The sequences identified from the color histogram are plotted as solid lines in Figure~\ref{fig:cmd_data}, and the sequences identified by eye are drawn as dotted lines. The sequences are in good agreement with each other. For the remainder of the paper, we adopt the sequences selected with the aid of stellar histograms and interpolated using a quadratic function to generate our empirical sequences.

The measurement of the HeB sequences can be made down to a lower luminosity limit where the sequences begin to blend with other features in an optical CMD. This lower luminosity limit depends on both the underlying structure in the CMD and the photometric errors. Specifically, the BHeB stars begin to merge with the red clump stars at fainter magnitudes while the RHeB stars blend with RGB stars. For example, for ESO~154$-$023 and NGC~2366 (Figure~\ref{fig:cmd_data}) the BHeB stars begin to merge into the red clump at $\sim$ M$_V \sim -3$, $V-I$ $\sim0.3$, and M$_V \sim -2$, $V-I$ $\sim0.4$, respectively. Similarly, the RHeB begin to merge with the RGB stars at M$_V \sim -4$, $V-I$ $\sim0.8$ and M$_V \sim -3.8$, $V-I$ $\sim1$ respectively in the same galaxies. Since the position of the RGB stars, HeB stars, and the red clump depends on the metallicity and the SFH \citep{Cole1998}, the lowest luminosity at which one can select HeBs from a CMD must be determined on a case-by-case basis. 

We use synthetic stellar populations to determine the luminosity and color for unambiguously separating HeB stars from other stars in each galaxy. We generated synthetic CMDs using the SFHs derived in \citet[][and references therein]{McQuinn2010b}, assumed a Salpeter IMF, and applied artificial star tests to mimic the uncertainties and completeness in the observed CMDs. Thus, the synthetic CMDs are close replicas of the observed CMDs, with the advantage that each star has a known age. We selected by eye the stars in the BHeB region, extending to depths expected to contain red clump stars, and stars in the RHeB region, extending to depths expected to contain RGB stars. The age distribution of these stars as a function of magnitude reveals the magnitude at which the young HeB stars become mixed with other stars. We define the lower luminosity limit for selecting HeB stars in each galaxy by the luminosity at which the young HeB stars become contaminated by $\sim5$\% of MS or RGB stars.

\begin{figure}
\centering
\subfigure{
	\includegraphics[scale=0.34]{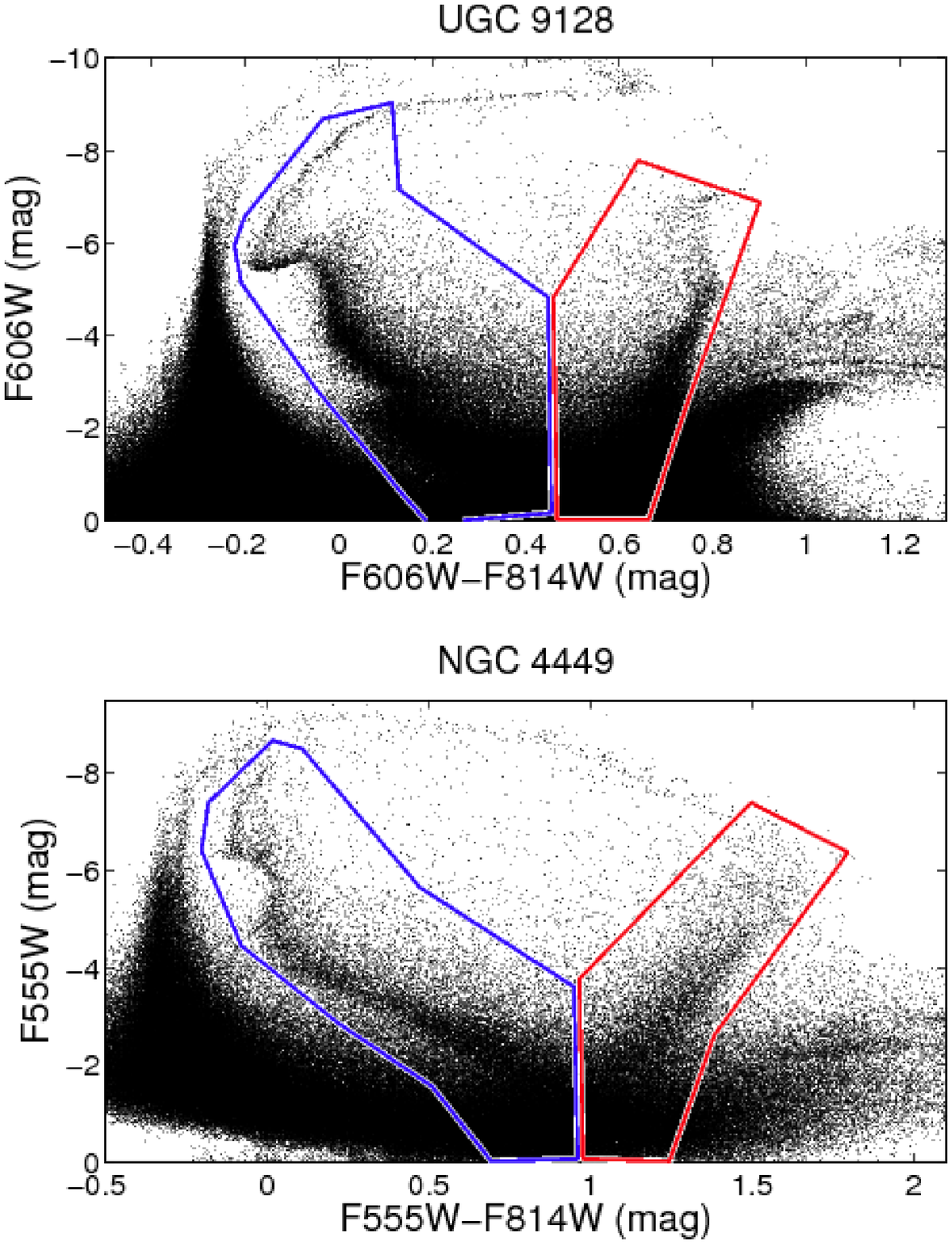}
}
\subfigure{
	\includegraphics[scale=0.25]{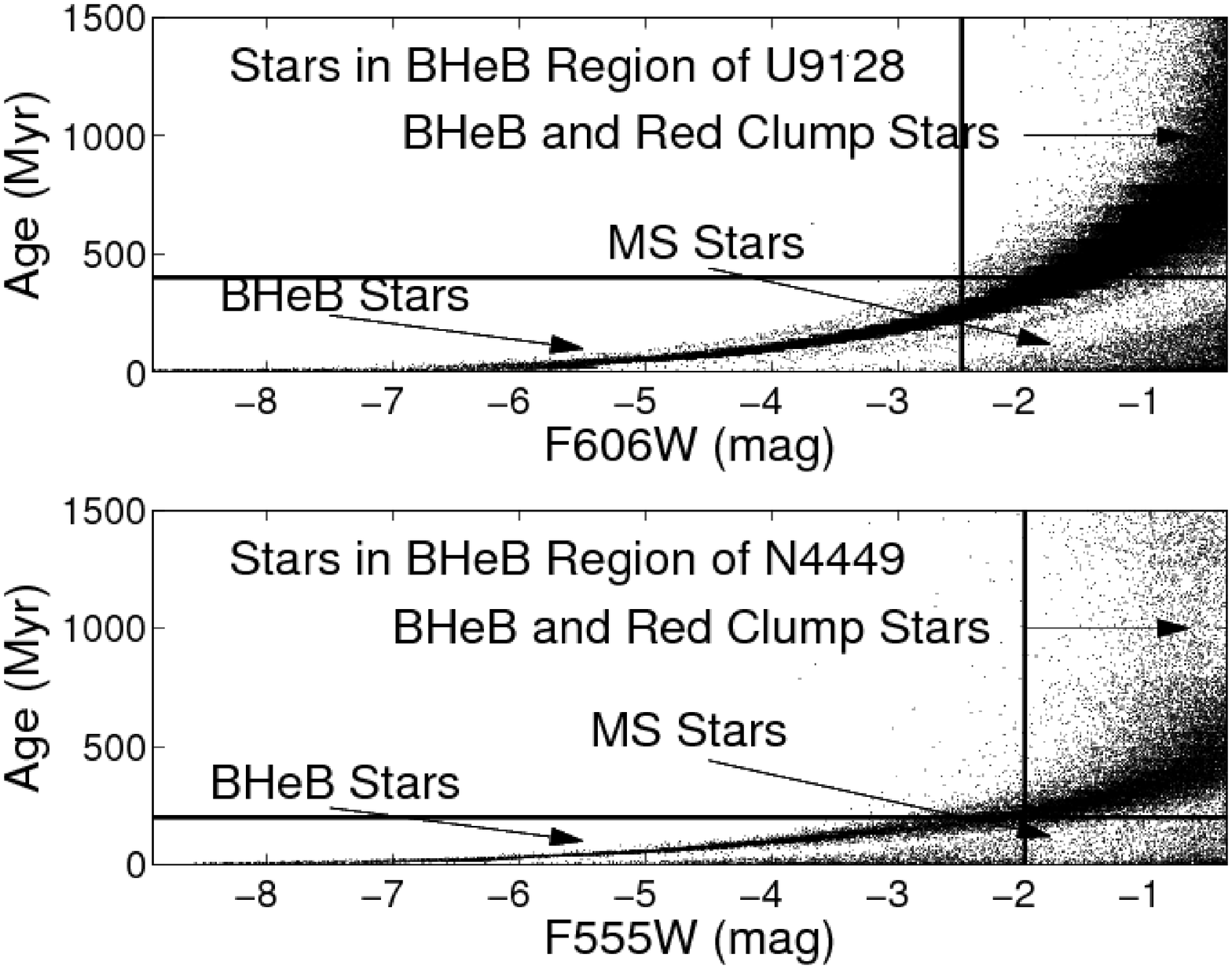}
}
\subfigure{
	\includegraphics[scale=0.25]{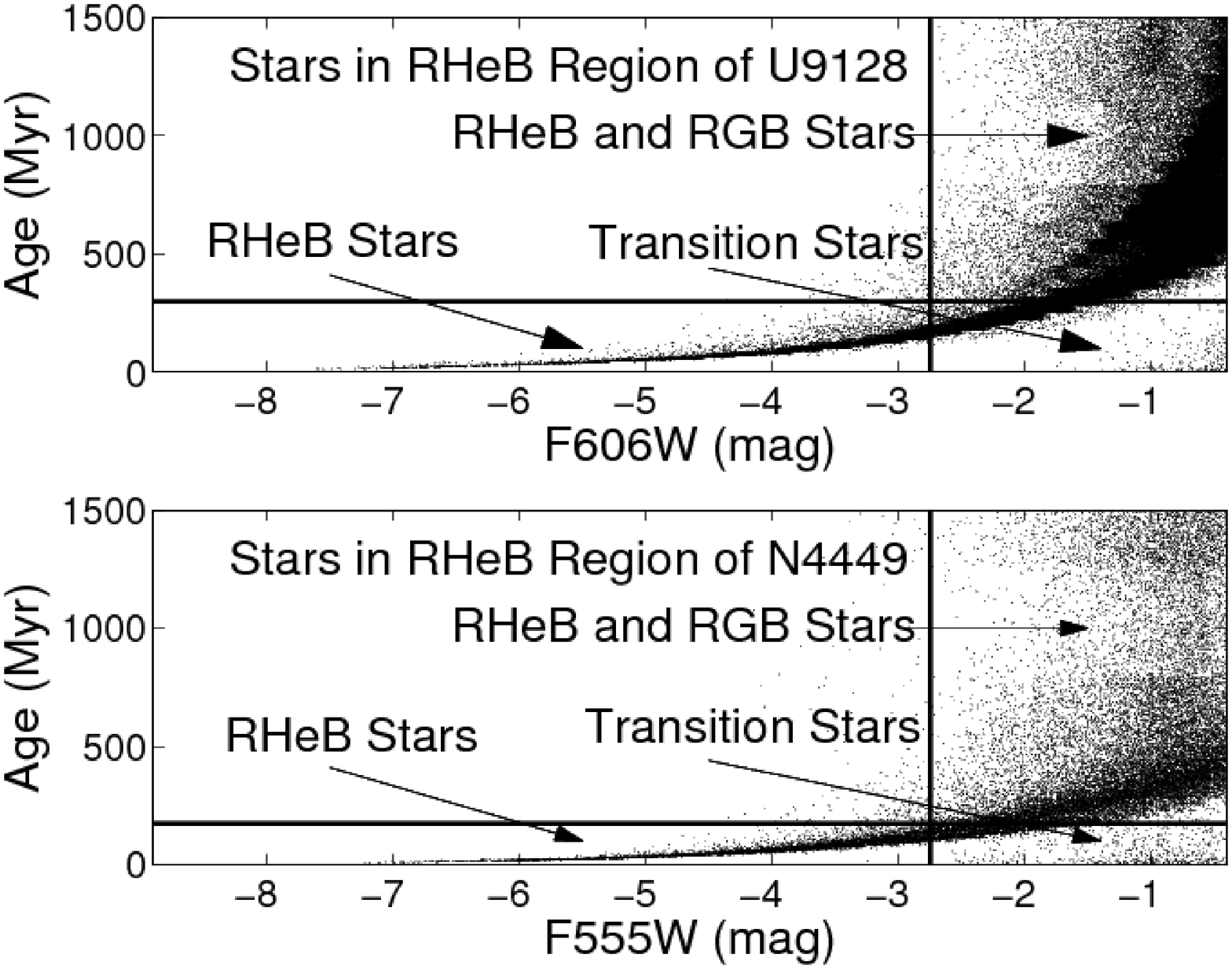}
}
\caption{Examples of the age distribution of stars in the blue and red HeB regions of an optical CMD. In the top two panels, the synthetic CMD of a lower and higher metallicity galaxy are shown (UGC~9128, $Z\sim 0.002$ and NGC~4449, $Z\sim0.009$). For illustrative purposes, we increased the SFRs at recent times to generate this CMD to ensure all luminosities are well-populated in the CMD. Outlined are the regions where the blue and red HeB stars are lie. Note there are more stars in the UGC~9128 simulated data as the SFRs used to simulate this galaxy were higher. In the middle two panels, the age distribution as a function of luminosity is shown for the stars selected in the BHeB region of each CMD. The bottom two panels show the distribution for the stars selected in the RHeB region. The vertical and horizontal lines in the middle and lower panels mark the limits in magnitudes and ages that the HeB stars can be uniquely separated in these CMDs with a contamination of order 5\% from non-HeB stars. Comparing the two systems, the BHeB can be separated for longer times at lower metallicity and for longer times than the RHeB stars at all metallicities. The RHeB stars becomes difficult to separate at younger ages in higher metallicity systems reducing their usefulness in tracing the recent SFHs in these systems.}
\label{fig:bheb_rheb_age}
\end{figure}

For illustrative purposes, in Figure~\ref{fig:bheb_rheb_age}, we present examples of the age distribution as a function of magnitude of candidate HeB stars for a lower metallicity galaxy (left panels$\colon Z\sim0.002$, UGC~9128) and for a higher metallicity galaxy (right panels$\colon Z\sim0.008$, NGC~4449). The synthetic CMDs generated for this demonstration have SFRs that increase at recent times to ensure all time bins are well-populated. From Figure~\ref{fig:bheb_rheb_age}, there is a one-to-one correspondence of age with magnitude at brighter luminosities, but once the HeB populations become mixed with older stars, the age distribution spreads dramatically. From the top panels in Figure~\ref{fig:bheb_rheb_age}, the BHeB stars blend with the older red clump populations at $\simeq400$ Myr and $\simeq300$ Myr for the lower and higher metallicity systems respectively. Note that the brighter luminosities of lower metallicity stars means the HeB stars can be traced farther back in time at lower metallicities.  At fainter magnitudes, the photometric uncertainties are higher, thus the BHeB region analyzed also contains some younger MS stars, as indicated in the figure. From the bottom panels of Figure~\ref{fig:bheb_rheb_age}, the RHeB stars blend with the older RGB population at $\ltsimeq300$ Myr and $\ltsimeq175$ Myr for the lower and higher metallicity systems. Comparing the age limits of the blue and red HeB stars for these two galaxies, Figure~\ref{fig:bheb_rheb_age} illustrates the more limited time interval probed when using RHeB stars as tracers of a galaxy's SFH compared with BHeB stars. At the faintest magnitudes analyzed, a few stars transitioning from the MS to RHeB stars are identified.

Note that the structure of the CMD changes at solar metallicities. At this metallicity, the RHeB stars are well-separated from the RGB stars at lower luminosities than in low metallicity systems, but begin to merge with BHeB stars after $\sim100$ Myr, as the ``blue loop'' is shortened to $\ltsimeq0.2$ mag in color. The merged HeB populations have a range in age of $\ltsimeq100$ Myr, thus the combined blue and red HeB stars can be used jointly to trace SF over timescales of order a few 100 Myr.

\subsection{The Empirically-Determined BHeB and RHeB Sequences}

\begin{figure}
\epsscale{1}
\plotone{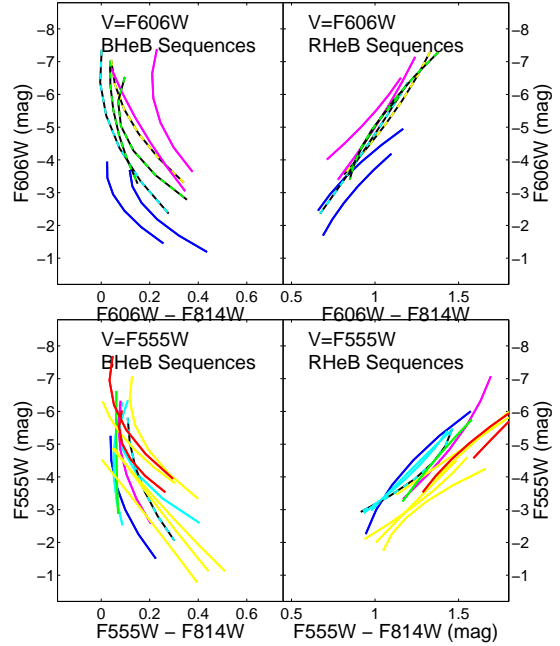}
\caption{Empirical BHeB and RHeB sequences for two filter combinations. The colors represent the different metallicities of the galaxies accorinding to the following key: blue is $Z=0.001$, cyan is $Z=0.002$, green is $Z=0.003$, yellow is $Z=0.005$, magenta is $Z=0.006$, red is $Z=0.009$. The colored lines were dashed with black for galaxies lacking \HII\ region abundances and whose metallicity was estimated from the SFHs. These sequences can be directly compared with the theoretically predicted sequences shown in Figure~\ref{fig:theory_iso}. }
\label{fig:emp_iso}
\end{figure}

The empirically determined red and blue HeB sequences are plotted in Figure~\ref{fig:emp_iso} for the galaxy sample. In the top and bottom panels, we present the sequences for the galaxies observed with the F555W and F814W filters and the F606W and F814W filters respectively. The sequences are color-coded by present-day metallicity values, with blue being the most metal poor ($0.0005 < Z <0.0015$), red being the most metal rich ($0.0080< Z <0.0095$), and the other colors spanning the intermediate metallicities. The colors of the lines match those used for the theoretical sequences in Figure~\ref{fig:theory_iso}. For metallicities derived from \HII\ region abundances, the lines in Figure~\ref{fig:emp_iso} are drawn as solid colors; for metallicities derived from the luminosity-metallicity relation, the lines are dashed with black.

\section{Comparing the Empirical and Theoretical HeB Results}\label{compare}

The empirically determined blue and red HeB sequences in Figure~\ref{fig:emp_iso} provide a general assessment of the location of the HeB stars for a given metallicity and filter combination and can be directly compared with the theoretically predicted HeB sequences in Figure~\ref{fig:theory_iso}. For easier comparison, we selected the sequences from one filter combination, F555W and F814W, and plot both the empirical and theoretical BHeB sequences together in Figure~\ref{fig:emp_theory_bheb}. Comparable RHeB sequences are plotted in Figure~\ref{fig:emp_theory_rheb}. We discuss each sequence separately followed by a comparison of the ratios of the number of stars in each sequence.

\subsection{Comparing BHeB Sequences}\label{compare_bheb}
\begin{figure}
\epsscale{1}
\plotone{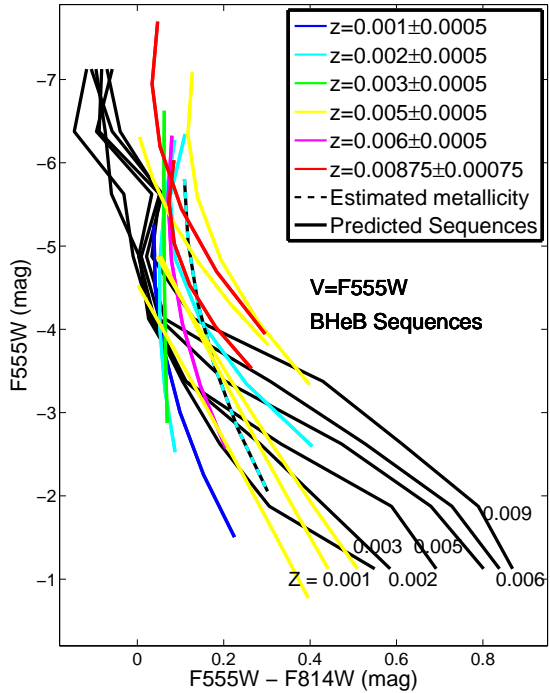}
\caption{The empirical BHeB sequences measured for the F555W and F814W filter observations from Figure~\ref{fig:emp_iso} plotted over the predicted sequences from Figure~\ref{fig:theory_iso}. The colors represents the metallicity of the galaxies sample as detailed in the figure legend. The predicted sequences are drawn in gray but span the same metallicity as the data. There is general agreement between the measured and predicted BHeB sequences. Differences exist with the theory predicted bluer (i.e., hotter) colors at brighter magnitudes than measured; the redder colors measured is likely due to extinction still present around these youngest BHeB stars.}
\label{fig:emp_theory_bheb}
\end{figure}

Figure~\ref{fig:emp_theory_bheb} shows that the empirically determined and theoretically predicted BHeB optical luminosities and colors \textit{agree} in two important characteristics. First, both sets of HeB sequences overlap in the CMDs and follow comparable trends between color and luminosities along an individual sequence (i.e., bluer colors at higher luminosities and redder colors at lower luminosities). Second, higher metallicity stars follow redder sequences in both the observed and synthetic CMDs.

There are also two main differences between the measured and predicted BHeB sequences. First, at lower luminosities (M$_V \gtsimeq -4$ mag) the slope of the observed sequences (i.e., $\Delta$V/$\Delta$(V$-$I)) is steeper than the slope of the predicted sequences, indicating that the lower mass BHeB are observed to be bluer than predicted. Although the photometric uncertainties are largest at these fainter magnitudes, they are uniformly distributed in color and would not bias the measurement of the empirical sequences towards bluer colors. This difference cannot be explained by photometric uncertainties or by extinction since the empirical sequences would be redder in color than observed. 

Second, we note an offset in color between the empirical and theoretical sequences at intermediate and higher luminosities (M$_V \ltsimeq -4$ mag). While both sets of sequences converge to a narrow range in color at high luminosities, the predicted sequences converge around F606W$-$F814W $\sim-0.10$ while the empirical sequences converge to redder colors of F606W$-$F814W $\sim0.05$, an offset in color of $\sim0.15$ mag. At these higher luminosities, the BHeB population is made up of very young (t$\ltsimeq50 $ Myr), massive (M$\gtsimeq 7$ \msun) stars. The redder colors of the empirically derived sequences may be attributable to dust co-located with these massive stars. To test this possibility, we added varying amounts of extinction to a set of synthetic CMDs and re-measured the location of the predicted BHeB sequences. The addition of A$_V\sim0.5$ mag of extinction applied to the young stellar population brighter than M$_V \sim4$ mag shifts the predicted sequences to redder colors, closely matching the empirical sequences. This amount of extinction is consistent with the estimated reddening found around OB stars in the Large Magellenic Cloud \citep{Harris1997}. Thus, while we cannot rule out the possibility that the predicted colors reflect a discrepancy in the models, the observed redder colors of the brightest BHeB stars may be attributable to extinction. However, the addition of  extinction would make the discrepancy worse for lower mass BHeB stars ($\ltsimeq 7$ \msun), if the dusty environment persists for longer timescales ($50-100$ Myr).

\subsection{Comparing RHeB Sequences }\label{compare_rheb}
\begin{figure}
\plotone{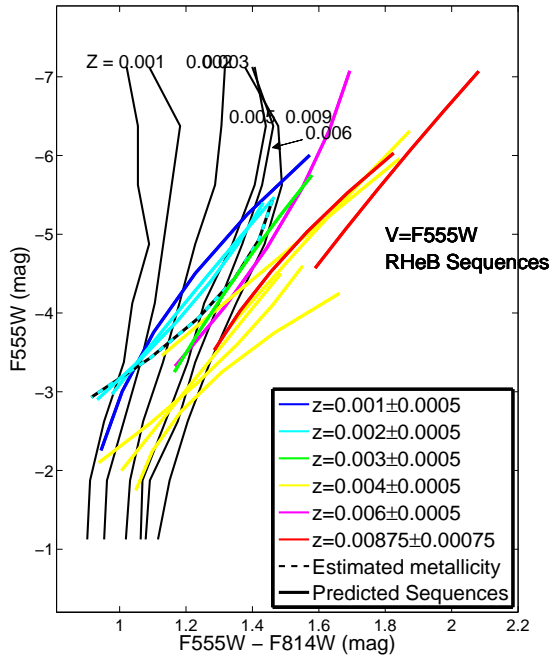}
\caption{The empirical RHeB sequences measured for the F555W and F814W filter observations from Figure~\ref{fig:emp_iso} plotted over the predicted sequences from Figure~\ref{fig:theory_iso}.The colors represents the metallicity of the galaxies sample as detailed in the figure legend. The predicted sequences are drawn in black and span the same metallicity as the data. There is general agreement between the measured and predicted BHeB sequences. Differences exist with the theory predicted bluer colors at brighter magnitudes than measured; however, unlike the BHeB stars, this color difference is too large to be mostly attributable to extinction around this very young population. In addition, the overall slope of the predicted sequences (i.e., $\Delta$F555W/$\Delta$(F555W$-$F814W)) is steeper than measured by a factor of 3.}
\label{fig:emp_theory_rheb}
\end{figure}

Like the BHeB sequences, the observed and theoretical RHeB sequences are in general agreement (Figure~\ref{fig:emp_theory_rheb}). The empirical and theoretical RHeB sequences both show a trend of higher metallicity sequences having redder colors.

However, there are greater differences between the measured and predicted RHeB sequences than the BHeB sequences (see \S\ref{compare_bheb}). First, the empirical RHeB sequences are redder, by as much as $\sim0.5$ mag, than the theoretical sequences at higher luminosities (M$_V \ltsimeq -4$ mag), for an equivalent metallicity. Unlike the 0.15 mag color offset in the BHeB sequences at bright magnitudes, this color difference cannot be fully attributable to differential extinction associated with young massive stars. Over three times the amount of extinction needed to explain the discrepancy in the BHeB sequences would be needed to account for the color shift in the RHeB sequences in spite of the stars being comparable masses and ages. Thus, unlike the difference between the observed and predicted color of the brightest BHeB stars, the bluer predicted colors of the brightest RHeB stars cannot be solely due to extinction.

Second, as seen in Figure~\ref{fig:emp_theory_rheb}, the slopes of the empirical sequences fall in a tight range ($\Delta$F555W/$\Delta$(F555W-F814W) is between $-3.9$ and $-7.0$ with an average value of $-4.9$). In contrast, the slopes of the theoretical sequences are a factor of $\sim3$ steeper ranging from $-13$ to $-25$ with an average of $-18$. The shallower slope of the empirical RHeB populations is thus a notable challenge for stellar models.  

\subsection{Comparing the Empirical and Theoretical B/R Ratios}

We can further test stellar evolution models by measuring the ratio of the blue to red HeB stars. For the synthetic CMDs, we select HeB stars within a width of $\pm0.1$ mag in color around each sequence (see Figure~\ref{fig:cmd_synth}). The small range in color is sufficient as the synthetic CMDs have no photometric uncertainty or extinction to widen the sequences in color. For the observed CMDs, the widths of the bins were selected on a case-by-case basis to best account for changing photometric uncertainties and extinction conditions (see Figure~\ref{fig:cmd_data} as an example of the photometric uncertainties as a function of magnitude).

\begin{figure}
\plotone{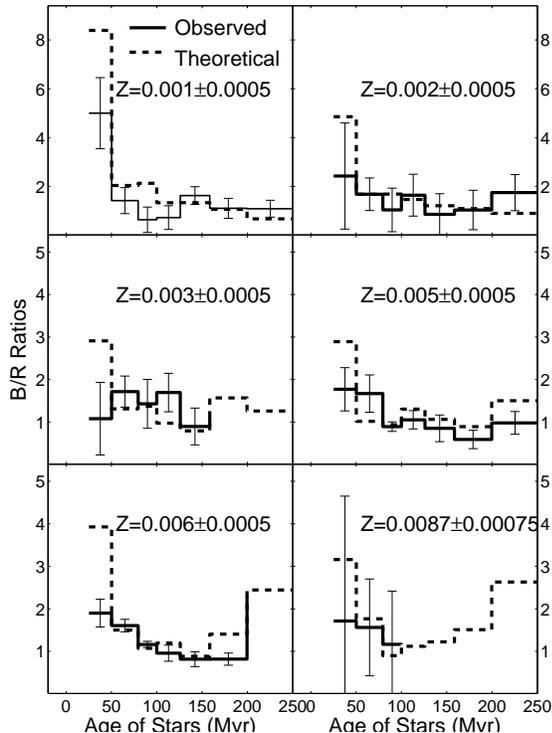}
\caption{Plotted are the average B/R ratios per metallicity bin for all galaxies in the sample (solid lines) and the B/R ratios per metallicity bin predicted from theoretical isochrones (dotted lines). The predicted B/R ratios are in general agreement with the measured B/R ratios. The most notable difference is the over-prediction of the B/R ratios for the youngest stars in the lowest metallicity studied confirming the results found by \citep{Dohm-Palmer2002} for Sextans~A, a galaxy of similar metallicity. The RHeB stars become difficult to isolate unambiguously from RGB stars at higher metallicities limiting the age over which the B/R ratio can be calculated. In Table~\ref{tab:ratios}, we list the number of blue and red HeB stars counted in each luminosity bins and the corresponding isochronal age.}
\label{fig:ratios}
\end{figure}

We measured the number of blue and red HeB stars as a function of magnitude. Using the stellar evolution isochrones, the magnitudes of both HeB phases were correlated with age. We matched the BHeB and RHeB stars of equal ages to calculate the ratio of B/R stars. In Table~\ref{tab:ratios}, we list the number of blue and red HeB stars as a function of luminosity with the corresponding age determined from theoretical isochrones at the appropriate metallicity. In Figure~\ref{fig:ratios}, we present the average B/R ratios measured per metallicity bin for the observed (solid lines) and theoretical CMDs (dashed lines). Although the plots extend to 250 Myr, we limit the observed ratio to ages where contamination from other evolutionary phases is less than 5\% as described in \S\ref{bheb}. The B/R ratios can be measured over the longest time periods at low metallicity, for which RHeB stars have brighter luminosities and are thus isolated in CMD space longer than higher metallicity RHeB stars. The exception to this is the $Z=0.005$ bin, which includes data from three fields in NGC~6822. These data are not only deeper due to the galaxy's proximity (such that the photometric uncertainties remain smaller at fainter magnitudes), but the SFH of NGC~6822 is such that the RHeB stars are distinguishable for longer times. In the highest metallicity bin of our sample ($Z=0.0087\pm0.00075 \simeq 0.5$~\zsun), the RHeB can only be robustly separated for $\ltsimeq 100$ Myr, providing a limited comparison with the predicted B/R ratio.

The uncertainties on the empirical B/R ratios in Figure~\ref{fig:ratios} include both Poisson errors and photometric uncertainties. The photometric uncertainties were calculated by measuring the difference between the B/R ratios from synthetic CMDs generated using artificial stars and from synthetic CMDs generated without applying artificial stars. At longer look-back times the uncertainty is dominated by the photometric uncertainties, because the larger number of fainter stars reduces the Poisson error. Note that in cases where the SFRs are rapidly changing, the B/R ratio is dependent on the SFH of a galaxy. However, given the stability of these ratios with time, any effect of the SFH is within our range of uncertainties. We tested the effects photometric crowding has on the B/R ratios by applying different crowding cuts to the photometry as stringent as 0.1 \citep[see][and the DOLPHOT photometry manual for a explanation of photometric crowding measurements]{Dolphin2000}. The B/R ratios were consistent across all crowding cuts, as the blue and red HeB stars are affected approximately equally by crowding. 

We find that the empirical and predicted B/R ratios are consistent, within the uncertainties, for nearly all time bins older than 50 Myr. Two sample Kolmogorov-Smirnov tests found the predicted and observed distribution of B/R ratios satisfied the null hypothesis of the observed and predicted ratios being drawn from the same distribution for each metallicity bin. For the youngest HeB stars studied (20 Myr $<$ Age $<$ 50 Myr), the predicted ratios are higher than we observe in all metallicities bins. A similar result was reported by \citet{Dohm-Palmer2002} for Sextans~A with a metallicity of $Z=0.001$. These authors find the theoretical isochrones over-predicted the B/R ratio by a factor of $\simeq2$.

\section{Discussion }\label{discussion}
Our result that the predicted HeB sequences and B/R ratios agree to first order with the observed sequences and B/R ratios confirms the usefulness of applying these isochrones to interpret observations of stellar populations. The discrepancies we report are likely to have subtle, second order effects on these interpretations. The reconstruction of SFHs leverages information from all stellar populations in a CMD, so that isolating the effects of model inadequacies in the HeB stars alone is complex. While measuring the impact of these discrepancies on SFH recovery is outside the scope of this work, we can make some initial speculations\footnote{We refer the reader to references listed in \S\ref{intro} for descriptions of deriving SFHs and \citet{McQuinn2010a} for an example of this application.}. We discuss the potential effects of the four discrepancies in turn.

First, the discrepancies found between the observed and predicted BHeB sequences for M$_V\gtsimeq-4$ mag are likely due to factors external to the stellar evolution models and thus must be accounted for separately when applying isochrones. The youngest HeB stars are often still co-located with material from their nascent gas cloud whose extinction likely accounts for the redder than predicted colors observed for these stars. Derivations of the SFHs include additional $A_V = 0.5$ mag of differential extinction to the models for stars younger than 40 Myr with a linear ramp-down of extinction with age for stars between 40-100 Myr \citep{Dolphin2002}, thus fitting the observed colors of the youngest HeB stars more closely. Second, the bluer color of the lower luminosity BHeB stars is likely to have the largest impact of the discrepancies on the derived SFRs. Stars in this region of the CMD may be mis-identified by the models as MS stars which can have a range in age at a given luminosity instead of the one-to-one age-luminosity relation of HeB stars. The color differences in the RHeB stars are greater for all magnitudes studied (i.e., the slopes of the observed and predicted HeB sequences as measured by $\Delta$V/$\Delta$(V$-$I) do not match) which cannot be entirely due to extinction. We expect the effect of this color difference to be of small consequence in the derivation of the SFHs as the RHeB stars are well-isolated in an optical CMD above the RGB stars making confusion of this population with a different aged population unlikely. The mis-match in the slope of the RHeB sequences degrades the quality of the fit between the modeled and the observed stellar populations, but will likely have a negligible effect on the derived SFRs. Fourth, as the predicted B/R ratio for stars younger than $\sim50$ Myr is too high, the derived SFRs may be underestimated, although this effect is mitigated as the derived SFRs also depend on the generally well-populated MS population.

A number of parameters in the stellar evolution models could explain the discrepancies between the predicted and observed red and blue HeB sequences and the B/R ratios including the amount of internal mixing assumed (e.g., the effects of convective overshoot), $^{12}$C$(\alpha,\gamma)^{16}$O reaction rates, and the effects of stellar rotation and mass loss. Additionally, the stellar atmosphere models used to convert the surface luminosity and $T_{eff}$ to magnitudes and colors may affect the predicted luminosity and color of the HeB stars. While the stellar atmosphere models show good agreement to observed stellar populations \citep[e.g.,][and references therein]{Girardi2002, Girardi2008}, the atmosphere grids are incomplete for the brightest giants at lower metallicities and thus required extrapolation to reach brighter luminosities. This extrapolation could also explain our noted discrepancies, particularly the difference between the predicted and observed slopes of the RHeB sequences. Until the stellar evolution and atmospheric models are adjusted to describe the data with a higher degree of accuracy, it is premature to correct for these effects in the SFHs, particularly as the discrepancies are second order effects.

\section{Conclusions }\label{conc}
We find general agreement between the observed red and blue HeB sequences and the B/R ratio in a sample of nineteen, starburst dwarf galaxies and the sequences predicted by Padova stellar evolution isochrones. We also find four notable differences between the empirical results and theoretical predictions. 

\begin{itemize}
\item There is a difference in the color of HeB stars at brighter luminosities (M$_V \ltsimeq-4$ mag), where the observed red and blue HeB sequences are redder than predicted by theoretical isochrones. We attribute at least part of the difference to the presence of $A_V \sim 0.5$ mag of extinction in the environs of the young HeB stars. However, while this extinction can produce the color difference between the predicted and observed BHeB sequences, it cannot account for the larger color difference between the predicted and observed RHeB sequences for HeB stars of comparable age. We note that for the bluest stars, U-band observations would provide the most leverage for comparing observed versus predicted effective temperatures. 
\item We see evidence that the observed BHeB stars are bluer than predicted by theoretical isochrones at lower luminosities (M$_V \gtsimeq -4$ mag). This difference cannot be explained by photometric uncertainties or additional extinction. Note that our conclusion is tentative as our sample is limited by photometric depth at these lower luminosities. 
\item The slope of the observed RHeB stars is shallower than predicted by the isochrones by more than a factor of 3 across the magnitude range studied. 
\item The theoretical isochrones over-predict the B/R ratio for the most massive stars at all metallicities studied, confirming the findings reported in \citet{Dohm-Palmer2002} for Sextans~A ($Z=0.001$). 
\end{itemize}

Additionally, the measurement of the lower luminosity limit for unambiguously separating RHeB stars in optical CMDs shows that the RHeB stars are difficult to separate from RGB stars in the highest metallicity bin of our sample ($Z\sim 0.5$~\zsun) at ages as young as 100 Myr, reducing the effectiveness of using RHeB stars as tracers of a galaxy's SFH past $\sim 100$ Myr in higher metallicity systems. Further, we note the importance of studying a larger sample of galaxies containing substantial blue and red HeB populations, low extinction, spanning a range of metallicities. Future analysis is warranted comparing different stellar evolution models with these results.

\section{Acknowledgments}
Support for this work was provided by NASA through a ROSES grant (No. NNX10AD57G). The authors are grateful to Philip Rosenfield for providing valuable comments on the stellar evolution models of HeB stars. E.~D.~S. is grateful for partial support from the University of Minnesota. K.~B.~W.~M. gratefully acknowledges Matthew, Cole, and Carling for their support. We are grateful to the anonymous referee for her/his helpful comments and suggestions.

{\it Facilities:} \facility{Hubble Space Telescope}


\begin{turnpage}
\begin{deluxetable}{lcccccccc}
\tabletypesize{\tiny}
\tablewidth{0pt}
\tablecaption{Galaxy Sample Extinction and Metallicity Values \label{tab:galaxies}}
\tablecolumns{9}
\tablehead{
\colhead{}				&
\colhead{HST V-band}			&
\colhead{$A_V$ (mag)}			&
\colhead{$A_V$ (mag)}			&
\colhead{$A_{814}$ (mag)}		&
\colhead{$A_{800}$ (mag)}		&
\colhead{Oxygen}			&
\colhead{}				&
\colhead{Equivalent}			\\
\colhead{Galaxy}			&
\colhead{filter}			&
\colhead{\citet{McQuinn2010b}}		&
\colhead{\citet{Schlegel1998}}		&
\colhead{Inferred}			&
\colhead{\citet{Schlegel1998}}		&
\colhead{Abundance}			&
\colhead{Source}			&
\colhead{Isochrone}			\\
\colhead{(1)}				&
\colhead{(2)}				&
\colhead{(3)}				&
\colhead{(4)}				&
\colhead{(5)}				&
\colhead{(6)}				&
\colhead{(7)}				&
\colhead{(8)}				&
\colhead{(9)}
}
\startdata
Antlia Dwarf	& F606W	& 0.20$\pm.04$ 			& 0.21 	& 0.15	& 0.15	  & 7.39  & 1 & 0.001 \\
UGC 9128	& F606W	& 0.20$\pm.04$ 			& 0.07 	& 0.15	& 0.05	  & 7.74  & 2 & 0.002  \\
UGC 4483	& F555W & 0.10$\pm.04$			& 0.11	& 0.07	& 0.07	  & 7.50  & 3 & 0.001  \\
NGC 4163	& F606W	& 0.00$\pm.04$ 			& 0.05 	& 0.00	& 0.04	  & 7.69  & 4 & 0.002   \\
UGC 6456	& F555W & 0.05$\pm.04$ 			& 0.12	& 0.04	& 0.07	  & 7.64  & 5 & 0.002  \\
NGC 6789	& F555W & 0.15$\pm.04$ 			& 0.23	& 0.11	& 0.14	  & 7.77  & 4 & 0.002   \\
NGC 4068	& F606W	& 0.00$\pm.03$			& 0.06 	& 0.00	& 0.04	  & 7.84  & 4 & 0.003   \\
DDO 165	 	& F555W & 0.10$\pm.04$			& 0.08	& 0.07	& 0.05	  & 7.80  & 6 & 0.002 	\\
IC 4662		& F606W	& 0.50$\pm.04$\tablenotemark{a} & 0.19 	& 0.36 	& 0.14 	  & 8.17  & 7 & 0.006  \\
ESO 154-023	& F606W	& 0.10$\pm.04$ 			& 0.05 	& 0.07	& 0.03	  & 8.01  & 4 & 0.004   \\
NGC 2366	& F555W & 0.20$\pm.04$\tablenotemark{a} & 0.12	& 0.15	& 0.07	  & 8.19  & 8 & 0.006 \\
NGC 625		& F555W & 0.00$\pm.04$			& 0.06	& 0.00	& 0.03	  & 8.10  & 9 & 0.005  \\
NGC 784		& F606W	& 0.10$\pm.04$			& 0.16	& 0.07	& 0.12	  & 8.05  & 4 & 0.004   \\
Ho II	 	& F555W & 0.10$\pm.04$			& 0.11	& 0.07	& 0.06	  & 7.92  & 6 & 0.003 \\
NGC 5253	& F555W	& 0.20$\pm.04$\tablenotemark{a} & 0.19	& 0.15	& 0.11	  & 8.10  & 10& 0.005  \\
NGC 6822	& F555W & 1.05$\pm0.1$\tablenotemark{a} & 0.78	& 0.76	& 0.46	  & 8.11  & 11& 0.005 \\
NGC 4214	& F555W & 0.15$\pm.04$			& 0.07	& 0.11	& 0.04	  & 8.38  & 12& 0.009  \\
NGC 1569	& F606W	& 1.80$\pm0.1$\tablenotemark{a} & 1.87  & 1.31 	& 1.36 	  & 8.19  & 13& 0.006  \\
NGC 4449	& F555W	& 0.25$\pm.04$\tablenotemark{a} & 0.06	& 0.18	& 0.04	  & 8.21  & 14& 0.008 \\
\enddata

\tablecomments{Col. (3) Estimated A$_V$ includes both foreground and internal extinction whereas extinction estimates in Col. (4) from \citet{Schlegel1998} measure only foreground extinction. Col (5) is the extinction inferred at an I central wavelength of 814 nm based on the A$_V$ value from Col. (3), assuming a reddening law of \citet{Cardelli1989} and calculated with the formulism by \citet{McCall2004}.  Col.(7) In 14 galaxies, oxygen abundances measurements were taken from the literature; in 5 galaxies oxygen abundances were estimated from the B-band luminosity-metallicity relation as noted in the Col.(8). All values assume a solar 12+[O/H] value of 8.69 \citep{Asplund2009}. Col. (9) Abundance measurements from Col. (7) converted to isochronal values.}
\tablenotetext{a}{We report the highest $A_V$ of total extinction measured for these galaxies. Lower levels of extinction for measured for regions of lower surface brightness in these systems.}

\tablerefs{(1) \citet{Piersimoni1999}; (2) \citet{vanZee1997}; (3) \citet{Skillman1994}; (4) L-Z relation; \citet{Zaritsky1994, Tremonti2004, Lee2006}; (5) \citet{Croxall2009}; (6) \citet{Hidalgo2001a}; (7) \citet{Roy1996}; (8) \citet{Skillman2003}; (9) \citet{Kobulnicky1997a}; (10) \citet{Hidalgo2001a}; (11) \citet{Kobulnicky1996}; (12) \citet{Kobulnicky1997b}; (13)\citet{Skillman1989}}
\end{deluxetable}
\end{turnpage}


\begin{deluxetable}{lcc}
\tabletypesize{\scriptsize}
\tablewidth{0pt}
\tablecaption{Instrument and Filter Combinations \label{tab:instr_filter}}
\tablecolumns{3}
\tablehead{
\colhead{HST Instrument}	&
\colhead{V band Filter}	&
\colhead{I band Filter}
}
\startdata
ACS WFC	&	F606W	&	F814W	\\
ACS WFC	&	F555W	&	F814W	\\
WFPC2	&	F555W	&	F814W	\\
\enddata

\tablecomments{The observations were obtained using two HST imaging instruments and span three difference instrument-filter combinations. The theorectical isochrones for the ACS and WFPC2 instruments using the same filter bandpasses were indistinguishable from one another. Thus, we present only the results from each of the two unique filter combinations. Additionally, although our observations were not obtained with the bluer F475W filter, we include the predictions for the filter combination of F475W and F814W for comparison.}
\end{deluxetable}

\clearpage
\LongTables
\begin{deluxetable}{lccrccr}
\tabletypesize{\tiny}
\tablewidth{0pt}
\tablecaption{Number of Blue and Red HeB stars as a Function of Age and Luminosity\label{tab:ratios}}
\tablecolumns{7}
\tablehead{
\colhead{Log}				&
\colhead{$M_V$}				&
\colhead{$M_{F814W}$}			&
\colhead{No. BHeB}			&
\colhead{$M_V$}				&
\colhead{$M_{F814W}$}			&
\colhead{No. RHeB}			\\
\colhead{Age}				&
\colhead{(mag)}				&
\colhead{(mag)}				&
\colhead{Stars}				&
\colhead{(mag)}				&
\colhead{(mag)}				&
\colhead{Stars}				
}
\startdata

\sidehead{ NGC~784, V=F606W }
7.4 & -6.20 & -6.18 &    36 & -6.24 & -7.27 &    60 \\
7.5 & -6.13 & -6.18 &     4 & -5.89 & -6.91 &    42 \\
7.6 & -5.71 & -5.77 &    58 & -5.48 & -6.47 &    63 \\
7.7 & -5.13 & -5.12 &   168 & -5.10 & -6.05 &   123 \\
7.8 & -4.61 & -4.60 &   267 & -4.69 & -5.63 &   158 \\
7.9 & -4.29 & -4.29 &   269 & -4.24 & -5.15 &   257 \\
8.0 & -3.89 & -3.91 &   483 & -3.86 & -4.73 &   400 \\
8.1 & -3.51 & -3.58 &   785 & -3.41 & -4.26 &   914 \\
8.2 & -3.17 & -3.33 &  1176 & -2.99 & -3.82 &  1795 \\
\sidehead{ IC~4662, V=F606W }
7.4 & -6.24 & -6.25 &    35 & -6.29 & -7.35 &    56 \\
7.5 & -6.09 & -6.13 &    11 & -5.93 & -6.99 &    30 \\
7.6 & -5.68 & -5.73 &    33 & -5.53 & -6.54 &    62 \\
7.7 & -5.19 & -5.21 &    80 & -5.06 & -6.04 &    84 \\
7.8 & -4.71 & -4.72 &   124 & -4.74 & -5.71 &    76 \\
7.9 & -4.37 & -4.40 &   137 & -4.32 & -5.26 &   131 \\
8.0 & -3.95 & -4.01 &   259 & -3.84 & -4.74 &   260 \\
8.1 & -3.62 & -3.75 &   307 & -3.47 & -4.34 &   380 \\
8.2 & -3.17 & -3.41 &   605 & -2.97 & -3.82 &   799 \\
8.3 & -2.69 & -3.05 &   828 & -2.79 & -3.63 &   372 \\
8.4 & -2.18 & -2.67 &   NaN & -2.45 & -3.26 &   636 \\
\sidehead{ Antlia, V=F606W }
7.4 & -5.65 & -5.58 &     0 & -6.50 & -7.29 &     0 \\
7.5 & -5.54 & -5.51 &     0 & -6.10 & -6.90 &     0 \\
7.6 & -5.28 & -5.26 &     0 & -5.62 & -6.43 &     0 \\
7.7 & -4.99 & -4.98 &     0 & -5.11 & -5.94 &     0 \\
7.8 & -4.60 & -4.60 &     0 & -4.66 & -5.46 &     1 \\
7.9 & -4.32 & -4.33 &     1 & -4.18 & -4.95 &     0 \\
8.0 & -3.93 & -3.96 &     1 & -3.75 & -4.50 &     3 \\
8.1 & -3.65 & -3.71 &     2 & -3.24 & -3.98 &     6 \\
8.2 & -3.36 & -3.46 &     2 & -2.88 & -3.60 &     7 \\
8.3 & -2.93 & -3.07 &    10 & -2.51 & -3.23 &    11 \\
8.4 & -2.55 & -2.70 &    27 & -2.17 & -2.87 &    27 \\
8.5 & -2.21 & -2.33 &    23 & -1.83 & -2.53 &    53 \\
8.6 & -1.78 & -1.99 &    29 & -1.53 & -2.22 &    32 \\
8.7 & -1.33 & -1.66 &    72 & -1.33 & -2.03 &    12 \\
\sidehead{ UGC~9128, V=F606W }
7.4 & -5.89 & -5.85 &     0 & -6.45 & -7.38 &     0 \\
7.5 & -5.76 & -5.75 &     0 & -6.01 & -6.93 &     0 \\
7.6 & -5.64 & -5.66 &     1 & -5.59 & -6.50 &     0 \\
7.7 & -5.06 & -5.05 &     3 & -5.10 & -6.01 &     2 \\
7.8 & -4.60 & -4.60 &    11 & -4.71 & -5.59 &    13 \\
7.9 & -4.24 & -4.25 &     4 & -4.21 & -5.06 &    24 \\
8.0 & -3.98 & -4.00 &    23 & -3.83 & -4.66 &    21 \\
8.1 & -3.60 & -3.65 &    56 & -3.57 & -4.39 &    20 \\
8.2 & -3.19 & -3.31 &   152 & -3.05 & -3.84 &    85 \\
8.3 & -2.88 & -3.05 &   117 & -2.64 & -3.41 &   138 \\
8.4 & -2.44 & -2.73 &   160 & -2.28 & -3.04 &    99 \\
8.5 & -2.01 & -2.35 &   313 & -1.94 & -2.69 &    72 \\
8.6 & -1.62 & -2.02 &   385 & -1.72 & -2.47 &    35 \\
8.7 & -1.20 & -1.67 &   464 & -1.45 & -2.22 &     0 \\
\sidehead{ NGC~4163, V=F606W }
7.4 & -5.94 & -5.88 &     3 & -6.40 & -7.30 &     8 \\
7.5 & -5.77 & -5.75 &     2 & -6.03 & -6.91 &     7 \\
7.6 & -5.57 & -5.58 &     2 & -5.60 & -6.48 &     8 \\
7.7 & -5.08 & -5.06 &    14 & -5.06 & -5.93 &    10 \\
7.8 & -4.64 & -4.62 &    23 & -4.70 & -5.55 &    11 \\
7.9 & -4.33 & -4.33 &    18 & -4.28 & -5.11 &    33 \\
8.0 & -3.91 & -3.91 &    40 & -3.89 & -4.69 &    35 \\
8.1 & -3.60 & -3.63 &    53 & -3.43 & -4.21 &    75 \\
8.2 & -3.18 & -3.26 &    44 & -2.99 & -3.76 &    53 \\
8.3 & -2.81 & -2.97 &   120 & -2.58 & -3.33 &   142 \\
8.4 & -2.43 & -2.68 &   159 & -2.26 & -3.00 &   126 \\
8.5 & -2.04 & -2.33 &   355 & -1.99 & -2.71 &   224 \\
\sidehead{ NGC~4068, V=F606W }
7.4 & -6.04 & -5.99 &    33 & -6.37 & -7.32 &    29 \\
7.5 & -5.99 & -6.00 &     1 & -5.96 & -6.90 &    17 \\
7.6 & -5.75 & -5.78 &    15 & -5.51 & -6.44 &    25 \\
7.7 & -5.13 & -5.11 &    68 & -5.16 & -6.06 &    36 \\
7.8 & -4.61 & -4.60 &   110 & -4.66 & -5.54 &    77 \\
7.9 & -4.21 & -4.21 &   195 & -4.23 & -5.09 &   136 \\
8.0 & -3.86 & -3.86 &   216 & -3.96 & -4.82 &   109 \\
8.1 & -3.56 & -3.60 &   274 & -3.47 & -4.29 &   361 \\
8.2 & -3.14 & -3.26 &   442 & -3.08 & -3.89 &   490 \\
8.3 & -2.79 & -2.99 &   499 & -2.65 & -3.43 &   529 \\
\sidehead{ ESO~154$-$023, V=F606W }
7.4 & -6.29 & -6.28 &    42 & -6.30 & -7.29 &    69 \\
7.5 & -6.10 & -6.13 &    18 & -5.99 & -6.97 &    33 \\
7.6 & -5.79 & -5.84 &    51 & -5.59 & -6.54 &    58 \\
7.7 & -5.16 & -5.15 &   174 & -5.10 & -6.03 &   125 \\
7.8 & -4.65 & -4.64 &   304 & -4.71 & -5.61 &   148 \\
7.9 & -4.29 & -4.28 &   317 & -4.26 & -5.13 &   286 \\
8.0 & -3.86 & -3.86 &   641 & -3.87 & -4.72 &   402 \\
8.1 & -3.53 & -3.56 &   631 & -3.51 & -4.36 &   611 \\
8.2 & -3.10 & -3.19 &  1089 & -3.10 & -3.91 &  1233 \\
8.3 & -2.77 & -2.98 &   NaN & -2.80 & -3.60 &  1142 \\
\sidehead{ NGC~1569, V=F606W }
7.4 & -6.04 & -6.16 &   660 & -6.26 & -7.42 &   819 \\
7.5 & -5.95 & -6.12 &    65 & -5.79 & -6.94 &   385 \\
7.6 & -5.54 & -5.71 &   384 & -5.42 & -6.53 &   369 \\
7.7 & -5.02 & -5.17 &   752 & -5.08 & -6.16 &   382 \\
7.8 & -4.52 & -4.67 &  1058 & -4.65 & -5.69 &   741 \\
7.9 & -4.27 & -4.45 &   604 & -4.20 & -5.23 &  1189 \\
8.0 & -3.86 & -4.06 &   961 & -3.82 & -4.84 &  1323 \\
\sidehead{ UGC~4483, V=F555W }
7.4 & -5.64 & -5.54 &     5 & -6.15 & -7.25 &     1 \\
7.5 & -5.54 & -5.50 &     1 & -5.71 & -6.81 &     0 \\
7.6 & -5.33 & -5.30 &     3 & -5.25 & -6.35 &     1 \\
7.7 & -4.96 & -4.94 &     4 & -4.80 & -5.90 &     3 \\
7.8 & -4.60 & -4.59 &    10 & -4.29 & -5.37 &     2 \\
7.9 & -4.26 & -4.27 &    11 & -3.84 & -4.88 &    10 \\
8.0 & -3.95 & -3.97 &    14 & -3.45 & -4.48 &    19 \\
8.1 & -3.61 & -3.67 &    26 & -3.05 & -4.05 &    15 \\
8.2 & -3.20 & -3.30 &    53 & -2.63 & -3.62 &    43 \\
8.3 & -2.86 & -3.00 &    66 & -2.31 & -3.27 &    44 \\
8.4 & -2.43 & -2.64 &   102 & -1.93 & -2.89 &    53 \\
8.5 & -2.06 & -2.26 &   156 & -1.59 & -2.55 &    51 \\
8.6 & -1.59 & -1.87 &   239 & -1.30 & -2.25 &    23 \\
8.7 & -1.16 & -1.57 &   332 & -1.11 & -2.06 &     0 \\
\sidehead{ UGC~6456, V=F555W }
7.4 & -5.78 & -5.70 &    42 & -6.13 & -7.31 &     7 \\
7.5 & -5.65 & -5.63 &     5 & -5.65 & -6.83 &     6 \\
7.6 & -5.48 & -5.49 &     9 & -5.23 & -6.42 &     6 \\
7.7 & -5.02 & -5.01 &    16 & -4.72 & -5.91 &    12 \\
7.8 & -4.57 & -4.55 &    27 & -4.27 & -5.44 &    11 \\
7.9 & -4.29 & -4.29 &    30 & -3.78 & -4.88 &    19 \\
8.0 & -3.92 & -3.94 &    47 & -3.45 & -4.53 &    15 \\
8.1 & -3.53 & -3.56 &    66 & -3.05 & -4.12 &    62 \\
8.2 & -3.17 & -3.31 &   105 & -2.60 & -3.65 &   126 \\
8.3 & -2.80 & -3.00 &   150 & -2.29 & -3.33 &   125 \\
8.4 & -2.37 & -2.64 &   153 & -1.92 & -2.96 &   NaN \\
\sidehead{ NGC~6789, V=F606W }
7.4 & -5.90 & -5.85 &     4 & -6.02 & -7.52 &     5 \\
7.5 & -5.81 & -5.80 &     0 & -5.64 & -7.08 &     4 \\
7.6 & -5.54 & -5.56 &     2 & -5.25 & -6.65 &     2 \\
7.7 & -5.05 & -5.06 &    13 & -4.75 & -6.10 &    14 \\
7.8 & -4.60 & -4.60 &    14 & -4.30 & -5.56 &    22 \\
7.9 & -4.23 & -4.23 &    15 & -3.87 & -5.07 &    26 \\
8.0 & -3.89 & -3.90 &    40 & -3.42 & -4.59 &    63 \\
8.1 & -3.55 & -3.58 &    36 & -3.05 & -4.15 &    87 \\
8.2 & -3.13 & -3.18 &    88 & -2.65 & -3.71 &   117 \\
8.3 & -2.71 & -2.92 &   116 & -2.28 & -3.33 &    41 \\
8.4 & -2.35 & -2.71 &   188 & -1.98 & -3.04 &     0 \\
8.5 & -1.86 & -2.28 &   261 & -1.64 & -2.70 &     0 \\
\sidehead{ DDO~165, V=F555W }
7.4 & -5.93 & -5.91 &    25 & -6.01 & -7.35 &    11 \\
7.5 & -5.85 & -5.87 &     7 & -5.62 & -6.94 &    12 \\
7.6 & -5.62 & -5.67 &    23 & -5.13 & -6.42 &    43 \\
7.7 & -5.02 & -5.04 &   181 & -4.79 & -6.07 &    55 \\
7.8 & -4.57 & -4.58 &   299 & -4.29 & -5.53 &   190 \\
7.9 & -4.23 & -4.24 &   447 & -3.87 & -5.08 &   291 \\
8.0 & -3.84 & -3.85 &   633 & -3.39 & -4.56 &   521 \\
8.1 & -3.47 & -3.48 &   548 & -3.02 & -4.15 &   420 \\
8.2 & -3.08 & -3.14 &   818 & -2.65 & -3.76 &   416 \\
8.3 & -2.72 & -2.84 &  1049 & -2.31 & -3.39 &   441 \\
8.4 & -2.23 & -2.40 &  2782 & -1.93 & -3.00 &   NaN \\
\sidehead{ NGC~2366, V=F555W }
7.4 & -6.04 & -6.05 &    64 & -5.71 & -7.30 &    15 \\
7.5 & -5.96 & -5.97 &     7 & -5.30 & -6.91 &    21 \\
7.6 & -5.56 & -5.59 &    54 & -4.92 & -6.50 &    43 \\
7.7 & -5.02 & -5.03 &   150 & -4.57 & -6.12 &    79 \\
7.8 & -4.66 & -4.68 &   162 & -4.17 & -5.70 &   112 \\
7.9 & -4.23 & -4.26 &   334 & -3.82 & -5.34 &   173 \\
8.0 & -3.87 & -3.91 &   462 & -3.41 & -4.91 &   402 \\
8.1 & -3.48 & -3.57 &   939 & -2.97 & -4.46 &   928 \\
8.2 & -3.00 & -3.31 &  1866 & -2.57 & -4.05 &  1768 \\
8.3 & -2.43 & -2.85 &  4079 & -2.21 & -3.69 &  1100 \\
\sidehead{ NGC~625, V=F555W }
7.4 & -6.13 & -6.12 &    28 & -5.81 & -7.21 &     8 \\
7.5 & -5.99 & -6.02 &     5 & -5.42 & -6.80 &    19 \\
7.6 & -5.68 & -5.72 &     9 & -5.10 & -6.44 &    17 \\
7.7 & -5.05 & -5.03 &    43 & -4.74 & -6.04 &    32 \\
7.8 & -4.52 & -4.48 &    99 & -4.24 & -5.51 &    84 \\
7.9 & -4.24 & -4.20 &    90 & -3.89 & -5.12 &    92 \\
8.0 & -3.89 & -3.88 &   139 & -3.51 & -4.72 &   187 \\
8.1 & -3.48 & -3.52 &   219 & -3.17 & -4.34 &   222 \\
8.2 & -3.02 & -3.14 &   299 & -2.71 & -3.87 &   NaN \\
\sidehead{ Ho~II, V=F555W }
7.4 & -6.10 & -6.09 &    99 & -5.92 & -7.30 &    66 \\
7.5 & -6.06 & -6.05 &    12 & -5.58 & -6.95 &    35 \\
7.6 & -5.76 & -5.79 &    57 & -5.16 & -6.50 &    64 \\
7.7 & -5.14 & -5.15 &   225 & -4.75 & -6.06 &   120 \\
7.8 & -4.63 & -4.63 &   380 & -4.24 & -5.51 &   218 \\
7.9 & -4.30 & -4.31 &   376 & -3.92 & -5.15 &   215 \\
8.0 & -3.87 & -3.89 &   671 & -3.50 & -4.70 &   444 \\
8.1 & -3.53 & -3.59 &   717 & -3.07 & -4.21 &   792 \\
8.2 & -3.13 & -3.26 &  1456 & -2.72 & -3.86 &   921 \\
8.3 & -2.72 & -3.00 &  2367 & -2.37 & -3.48 &   NaN \\
\sidehead{ NGC~5253, V=F555W }
7.4 & -6.20 & -6.22 &   206 & -5.75 & -7.24 &   169 \\
7.5 & -6.07 & -6.10 &    33 & -5.37 & -6.88 &   103 \\
7.6 & -5.71 & -5.79 &   118 & -5.00 & -6.43 &   116 \\
7.7 & -5.05 & -5.08 &   392 & -4.61 & -6.02 &   170 \\
7.8 & -4.60 & -4.62 &   381 & -4.27 & -5.63 &   212 \\
7.9 & -4.21 & -4.25 &   493 & -3.81 & -5.13 &   467 \\
8.0 & -3.93 & -4.01 &   443 & -3.36 & -4.64 &   NaN \\
8.1 & -3.50 & -3.61 &   957 & -3.00 & -4.26 &   NaN \\
\sidehead{ NGC~6822(1),\tablenotemark{a} V=F555W }
7.4 & -6.07 & -6.20 &     0 & -5.81 & -7.31 &     0 \\
7.5 & -5.99 & -6.15 &     0 & -5.45 & -6.95 &     0 \\
7.6 & -5.64 & -5.79 &     0 & -5.10 & -6.54 &     0 \\
7.7 & -5.03 & -5.18 &     0 & -4.71 & -6.11 &     0 \\
7.8 & -4.58 & -4.72 &     0 & -4.24 & -5.62 &     0 \\
7.9 & -4.23 & -4.37 &     2 & -3.81 & -5.15 &     4 \\
8.0 & -3.81 & -3.97 &    16 & -3.47 & -4.79 &     9 \\
8.1 & -3.39 & -3.61 &    15 & -3.10 & -4.38 &    15 \\
8.2 & -2.99 & -3.29 &    19 & -2.62 & -3.87 &    24 \\
8.3 & -2.51 & -3.03 &    28 & -2.34 & -3.57 &    18 \\
8.4 & -2.04 & -2.72 &    30 & -2.06 & -3.27 &    35 \\
8.5 & -1.44 & -2.26 &    48 & -1.70 & -2.90 &    61 \\
8.6 & -0.96 & -1.84 &    51 & -1.41 & -2.60 &   100 \\
8.7 & -0.52 & -1.45 &    71 & -1.04 & -2.23 &   134 \\
\sidehead{ NGC~6822(2),\tablenotemark{a} V=F555W }
7.4 & -6.10 & -6.24 &     0 & -5.85 & -7.35 &     0 \\
7.5 & -6.02 & -6.17 &     0 & -5.50 & -6.99 &     0 \\
7.6 & -5.62 & -5.80 &     0 & -5.05 & -6.50 &     0 \\
7.7 & -5.05 & -5.21 &     0 & -4.68 & -6.08 &     0 \\
7.8 & -4.58 & -4.73 &     1 & -4.35 & -5.73 &     0 \\
7.9 & -4.18 & -4.33 &     5 & -3.86 & -5.21 &     0 \\
8.0 & -3.82 & -4.00 &    16 & -3.45 & -4.76 &    15 \\
8.1 & -3.48 & -3.71 &    18 & -3.03 & -4.29 &    30 \\
8.2 & -3.02 & -3.37 &    14 & -2.68 & -3.89 &    44 \\
8.3 & -2.51 & -3.04 &    34 & -2.34 & -3.55 &    48 \\
8.4 & -2.00 & -2.69 &    56 & -2.04 & -3.24 &    59 \\
8.5 & -1.49 & -2.31 &    90 & -1.69 & -2.86 &   148 \\
8.6 & -1.00 & -1.90 &   141 & -1.38 & -2.55 &   218 \\
8.7 & -0.48 & -1.40 &   261 & -1.13 & -2.30 &   158 \\
\sidehead{ NGC~6822(3),\tablenotemark{a} V=F555W }
7.4 & -6.16 & -6.26 &     0 & -5.85 & -7.33 &     0 \\
7.5 & -6.02 & -6.17 &     0 & -5.45 & -6.91 &     0 \\
7.6 & -5.70 & -5.85 &     0 & -5.06 & -6.47 &     0 \\
7.7 & -5.02 & -5.11 &     0 & -4.69 & -6.06 &     0 \\
7.8 & -4.58 & -4.68 &     0 & -4.27 & -5.60 &     0 \\
7.9 & -4.27 & -4.38 &     0 & -3.87 & -5.19 &     0 \\
8.0 & -3.87 & -4.01 &     2 & -3.45 & -4.75 &     2 \\
8.1 & -3.45 & -3.65 &     3 & -3.13 & -4.38 &     4 \\
8.2 & -3.05 & -3.36 &     6 & -2.72 & -3.92 &     9 \\
8.3 & -2.52 & -2.99 &    13 & -2.35 & -3.53 &    19 \\
8.4 & -1.92 & -2.56 &    20 & -1.98 & -3.13 &    23 \\
8.5 & -1.49 & -2.25 &    11 & -1.70 & -2.83 &    29 \\
8.6 & -1.05 & -1.91 &    15 & -1.39 & -2.55 &    65 \\
8.7 & -0.48 & -1.35 &    22 & -1.11 & -2.27 &    40 \\
\sidehead{ NGC~4214, V=F555W }
7.4 & -6.13 & -6.15 &   115 & -5.51 & -7.12 &    68 \\
7.5 & -5.93 & -5.98 &    26 & -5.14 & -6.72 &    84 \\
7.6 & -5.50 & -5.53 &    95 & -4.79 & -6.30 &   112 \\
7.7 & -5.06 & -5.10 &   134 & -4.46 & -5.92 &   109 \\
7.8 & -4.69 & -4.75 &   189 & -4.07 & -5.46 &   208 \\
7.9 & -4.27 & -4.39 &   323 & -3.76 & -5.12 &   249 \\
8.0 & -3.82 & -4.05 &   517 & -3.42 & -4.74 &   460 \\
8.1 & -3.28 & -3.69 &   756 & -2.94 & -4.21 &   NaN \\
\sidehead{ NGC~4449, V=F555W }
7.4 & -6.04 & -6.03 &  1009 & -5.54 & -7.19 &   581 \\
7.5 & -5.84 & -5.86 &   251 & -5.16 & -6.80 &   513 \\
7.6 & -5.47 & -5.50 &   638 & -4.82 & -6.39 &   616 \\
7.7 & -5.02 & -5.03 &  1261 & -4.52 & -6.04 &   665 \\
7.8 & -4.71 & -4.75 &  1348 & -4.18 & -5.67 &  1058 \\
7.9 & -4.30 & -4.38 &  2443 & -3.75 & -5.17 &  2363 \\
8.0 & -3.87 & -4.02 &  3858 & -3.41 & -4.81 &  3209 \\
8.1 & -3.47 & -3.79 &  5072 & -2.97 & -4.31 &   NaN \\

\enddata

\tablecomments{The number of BHeB and RHeB stars per V-band and I-band luminosity bin and the associated age derived from isochrones. The V-band filters for each galaxy are listed next to the galaxy names; the I-band filter is F814W for all galaxies. NaN is listed for luminosities reaching below the photometric limit one can unambiguously separate blue or red HeB stars.}

\tablenotetext{a}{The observations for NGC~6822 covered three fields of view.}

\end{deluxetable}


\begin{thebibliography}{}
\bibitem[Alongi et al.(1993)]{Alongi1993} Alongi, M., Bertelli, G., Bressan, A., Chiosi, C., Fagotto, F., Greggio, L., \& Nasi, E.\ 1993, \aaps, 97, 851 
\bibitem[Aparicio et al.(1996)]{Aparicio1996} Aparicio, A., Gallart, 
C., Chiosi, C., \& Bertelli, G.\ 1996, \apjl, 469, L97 
\bibitem[Aparicio \& Hidalgo(2009)]{Aparicio2009} Aparicio, A., \& Hidalgo, S.~L.\ 2009, \aj, 138, 558 
\bibitem[Asplund et al.(2009)]{Asplund2009} Asplund, M., Grevesse, N., Sauval, A.~J., \& Scott, P.\ 2009, \araa, 47, 481 
\bibitem[Austin et al.(1971)]{Austin1971} Austin, S.~M., Trentelman, G.~F., \& Kashy, E.\ 1971, \apjl, 163, L79 
\bibitem[Bertelli et al.(1985)]{Bertelli1985} Bertelli, G., Bressan, A.~G., \& Chiosi, C.\ 1985, \aap, 150, 33 
\bibitem[Bertelli et al.(1994)]{Bertelli1994} Bertelli, G., Bressan, A., Chiosi, C., Fagotto, F., \& Nasi, E.\ 1994, \aaps, 106, 275 
\bibitem[Boquien et al.(2009)]{Boquien2009} Boquien, M., et al.\ 
2009, \apj, 706, 553 
\bibitem[Bressan et al.(1993)]{Bressan1993} Bressan, A., Fagotto, F., Bertelli, G., \& Chiosi, C.\ 1993, \aaps, 100, 647 
\bibitem[Bressan et al.(1998)]{Bressan1998} Bressan, A., Granato, G.~L., \& Silva, L.\ 1998, \aap, 332, 135 
\bibitem[Bruzual \& Charlot(2003)]{Bruzual2003} Bruzual, G., \& Charlot, S.\ 2003, \mnras, 344, 1000 
\bibitem[Cardelli et al.(1989)]{Cardelli1989} Cardelli, J.~A., 
Clayton, G.~C., \& Mathis, J.~S.\ 1989, \apj, 345, 245 
\bibitem[Calzetti et al.(1994)]{Calzetti1994} Calzetti, D., Kinney, A.~L., \& Storchi-Bergmann, T.\ 1994, \apj, 429, 582 
\bibitem[Castelli \& Kurucz(2003)]{Castelli2003} Castelli, F. , \& Kurucz, R.~L. 2003, in IAU Symp. 210, Modelling of Stellar Atmospheres, eds.N. Piskunov , W. W. Weiss , \& D. F. Gray (San Francisco: ASP), p. A20
\bibitem[Chiosi et al.(1992)]{Chiosi1992} Chiosi, C., Bertelli, G., \& Bressan, A.\ 1992, \araa, 30, 235 
\bibitem[Cole(1998)]{Cole1998} Cole, A.~A.\ 1998, \apjl, 500, 
L137 
\bibitem[Croxall et al.(2009)]{Croxall2009} Croxall, K.~V., van Zee, L., Lee, H., Skillman, E.~D., Lee, J.~C., C{\^o}t{\'e}, S., Kennicutt, R.~C., \& Miller, B.~W.\ 2009, \apj, 705, 723 
\bibitem[Dohm-Palmer et al.(1997)]{Dohm-Palmer1997} Dohm-Palmer, R.~C., 
et al.\ 1997, \aj, 114, 2527 
\bibitem[Dohm-Palmer \& Skillman(2002)]{Dohm-Palmer2002} Dohm-Palmer, R.~C., \& Skillman, E.~D.\ 2002, \aj, 123, 1433 
\bibitem[Dolphin(2000)]{Dolphin2000} Dolphin, A.~E.\ 2000, \pasp, 
112, 1383 
\bibitem[Dolphin(2002)]{Dolphin2002} Dolphin, A.~E., 2002, \mnras, 332, 91
\bibitem[Fagotto et al.(1994a)]{Fagotto1994a} Fagotto, F., Bressan, A., Bertelli, G., \& Chiosi, C.\ 1994, \aaps, 104, 365 
\bibitem[Dotter et al.(2008)]{Dotter2008} Dotter, A., Chaboyer, B., Jevremovi{\'c}, D., Kostov, V., Baron, E., 
\& Ferguson, J.~W.\ 2008, \apjs, 178, 89 
\bibitem[Fagotto et al.(1994b)]{Fagotto1994b} Fagotto, F., Bressan, A., Bertelli, G., \& Chiosi, C.\ 1994, \aaps, 105, 39 
\bibitem[Ford et al.(1998)]{Ford1998} Ford, H.~C., et al.\ 1998, 
\procspie, 3356, 234 
\bibitem[Gallart et al.(2005)]{Gallart2005} Gallart, C., Zoccali, M., \& Aparicio, A.\ 2005, \araa, 43, 387 
\bibitem[Girardi et al.(2000)]{Girardi2000} Girardi, L., Bressan, A., Bertelli, G., \& Chiosi, C.\ 2000, \aaps, 141, 371 
\bibitem[Girardi et al.(2002)]{Girardi2002} Girardi, L., Bertelli, G., Bressan, A., Chiosi, C., Groenewegen, M.~A.~T., Marigo, P., Salasnich, B., \& Weiss, A.\ 2002, \aap, 391, 195 
\bibitem[Girardi et al.(2004)]{Girardi2004} Girardi, L., Grebel, E.~K., Odenkirchen, M., \& Chiosi, C.\ 2004, \aap, 422, 205 
\bibitem[Girardi et al.(2008)]{Girardi2008} Girardi, L., et al.\ 2008, \pasp, 120, 583 
\bibitem[Groenewegen(2006)]{Groenewegen2006} Groenewegen, M.~A.~T.\ 2006, \aap, 448, 181 
\bibitem[Harris et al.(1997)]{Harris1997} Harris, J., Zaritsky, 
D., \& Thompson, I.\ 1997, \aj, 114, 1933 
\bibitem[Harris \& Zaritsky(2001)]{Harris2001} Harris, J., \& Zaritsky, D.\ 2001, \apjs, 136, 25 
\bibitem[Hayashi \& Cameron(1962)]{Hayashi1962a} Hayashi, C., \& Cameron, R.~C.\ 1962, \apj, 136, 166 
\bibitem[Hayashi et al.(1962)]{Hayashi1962b} Hayashi, C., H{\= o}shi, R., \& Sugimoto, D.\ 1962, Progress of Theoretical Physics Supplement, 22, 1 
\bibitem[Hidalgo-G{\'a}mez et al.(2001a)]{Hidalgo2001a} Hidalgo-G{\'a}mez, A.~M., Masegosa, J., \& Olofsson, K.\ 2001, \aap, 369, 797 
\bibitem[Hidalgo-G{\'a}mez et al.(2001b)]{Hidalgo2001b} Hidalgo-G{\'a}mez, A.~M., Olofsson, K., \& Masegosa, J.\ 2001, \aap, 367, 388 
\bibitem[Holtzman et al.(1995)]{Holtzman1995} Holtzman, J.~A., et 
al.\ 1995, \pasp, 107, 156 
\bibitem[Iben(1966)]{Iben1966} Iben, I., Jr.\ 1966, \apj, 143, 516
\& Rey, S.-C.\ 2009, \apj, 703, 614 
\bibitem[Kobulnicky \& Skillman(1996)]{Kobulnicky1996} Kobulnicky, H.~A., \& Skillman, E.~D.\ 1996, \apj, 471, 211 
\bibitem[Kobulnicky et al.(1997a)]{Kobulnicky1997a} Kobulnicky, H.~A., 
Skillman, E.~D., Roy, J.-R., Walsh, J.~R., \& Rosa, M.~R.\ 1997, \apj, 477, 679 
\bibitem[Kobulnicky \& Skillman(1997b)]{Kobulnicky1997b} Kobulnicky, H.~A., \& Skillman, E.~D.\ 1997, \apj, 489, 636 
\bibitem[Kurucz(1992)]{Kurucz} Kurucz, R.~L.\ 1992, in The Stellar Populations of Galaxies, ed. B. Barbuy, A. Renzini (Dordrecht, Kluwer), IAU Symp., 149, 225
\bibitem[Langer \& Maeder(1995)]{Langer1995} Langer, N., \& Maeder, A.\ 1995, \aap, 295, 685 
\bibitem[Ledoux(1947)]{Ledoux1947} Ledoux, P.\ 1947, \aj, 52, 155 
\bibitem[Lee et al.(2006)]{Lee2006} Lee, H., Skillman, E.~D., 
\& Venn, K.~A.\ 2006, \apj, 642, 813 
\bibitem[Leitherer et al.(2010)]{Leitherer2010} Leitherer, C., Ortiz 
Ot{\'a}lvaro, P.~A., Bresolin, F., Kudritzki, R.-P., Lo Faro, B., 
Pauldrach, A.~W.~A., Pettini, M., \& Rix, S.~A.\ 2010, \apjs, 189, 309 
\bibitem[Lejeune \& Schaerer(2001)]{Lejeune2001} Lejeune, T., \& Schaerer, D.\ 2001, \aap, 366, 538 
\bibitem[Maeder \& Meynet(2001)]{Maeder2001} Maeder, A., \& Meynet, G.\ 2001, \aap, 373, 555 
\bibitem[Marigo et al.(2008)]{Marigo2008} Marigo, P., Girardi, L., Bressan, A., Groenewegen, M.~A.~T., Silva, L., \& Granato, G.~L.\ 2008, \aap, 482, 883 
\bibitem[Meynet \& Maeder(1997)]{Meynet1997} Meynet, G., \& Maeder, A.\ 1997, \aap, 321, 465 
\bibitem[McCall(2004)]{McCall2004} McCall, M.~L.\ 2004, \aj, 128, 
2144 
\bibitem[McQuinn et al.(2009)]{McQuinn2009} McQuinn, K.~B.~W., 
Skillman, E.~D., Cannon, J.~M., Dalcanton, J.~J., Dolphin, A., Stark, D., 
\& Weisz, D.\ 2009, \apj, 695, 561 
\bibitem[McQuinn et al.(2010a)]{McQuinn2010a} McQuinn, K.~B.~W., Skillman, E.~D., Cannon, J.~M., Dalcanton, J.~J., Dolphin, A., Hidalgo-Rodriguez, S., Holtzman, J., Stark, D. et al., 2010, \apj, 721, 297
\bibitem[McQuinn et al.(2010b)]{McQuinn2010b} McQuinn, K.~B.~W., Skillman, E.~D., Cannon, J.~M., Dalcanton, J.~J., Dolphin, A., Hidalgo-Rodriguez, S., Holtzman, J., Stark, D. et al., 2010, \apj, in press
\bibitem[Piersimoni et al.(1999)]{Piersimoni1999} Piersimoni, A.~M., Bono, G., Castellani, M., Marconi, G., Cassisi, S., Buonanno, R., \& Nonino, M.\ 1999, \aap, 352, L63 
\bibitem[Pietrinferni et al.(2004)]{Pietrinferni2004} Pietrinferni, A., 
Cassisi, S., Salaris, M., \& Castelli, F.\ 2004, \apj, 612, 168 
\bibitem[Roy et al.(1996)]{Roy1996} Roy, J.-R., Belley, J., 
Dutil, Y., \& Martin, P.\ 1996, \apj, 460, 284 
\bibitem[Salpeter(1955)]{Salpeter1955} Salpeter, E.~E.\ 1955, \apj, 121, 161
\bibitem[Sakashita et al.(1959)]{Sakashita1959} Sakashita, S., 
{\^O}no, Y., \& Hayashi, C.\ 1959, Progress of Theoretical Physics, 21, 315 
\bibitem[Schlegel et al.(1998)]{Schlegel1998} Schlegel, D.~J., Finkbeiner, D.~P., \& Davis, M.\ 1998, \apj, 500, 525 
\bibitem[Skillman et al.(1989)]{Skillman1989} Skillman, E.~D., Kennicutt, R.~C., \& Hodge, P.~W.\ 1989, \apj, 347, 875 
\bibitem[Skillman et al.(1994)]{Skillman1994} Skillman, E.~D., 
Televich, R.~J., Kennicutt, R.~C., Jr., Garnett, D.~R., \& Terlevich, E.\ 1994, \apj, 431, 172 
\bibitem[Skillman et al.(2003)]{Skillman2003} Skillman, E.~D., 
C{\^o}t{\'e}, S., \& Miller, B.~W.\ 2003, \aj, 125, 610 
\bibitem[Stothers(1966)]{Stothers1966} Stothers, R.\ 1966, \apj, 143, 91 
\bibitem[Stothers \& Chin(1968)]{Stothers1968} Stothers, R., \& Chin, C.-W.\ 1968, \apj, 152, 225 
\bibitem[Stothers \& Chin(1975)]{Stothers1975} Stothers, R., \& Chin, C.-W.\ 1975, \apj, 198, 407 
\bibitem[Stothers \& Chin(1976)]{Stothers1976} Stothers, R., \& Chin, C.-W.\ 1976, \apj, 204, 472 
\bibitem[Stothers \& Chin(1991)]{Stothers1991} Stothers, R.~B., \& Chin, C.-W.\ 1991, \apj, 374, 288 
\bibitem[Stothers \& Chin(1992)]{Stothers1992} Stothers, R.~B., \& Chin, C.-W.\ 1992, \apj, 390, 136 
\bibitem[Tremonti et al.(2004)]{Tremonti2004} Tremonti, C.~A., et 
al.\ 2004, \apj, 613, 898 
\bibitem[Tolstoy \& Saha(1996)]{Tolstoy1996} Tolstoy, E., \& Saha, A.\ 1996, \apj, 462, 672 
\bibitem[VandenBerg et al.(2000)]{VandenBerg2000} VandenBerg, D.~A.,
Swenson, F.~J., Rogers, F.~J., Iglesias, C.~A.,
\& Alexander, D.~R.\ 2000, \apj, 532, 430
\bibitem[van Zee et al.(1997)]{vanZee1997} van Zee, L., Haynes, 
M.~P., \& Salzer, J.~J.\ 1997, \aj, 114, 2479 
\bibitem[Yi et al.(2001)]{Yi2001} Yi, S., Demarque, P., Kim, Y.-C., Lee, Y.-W., Ree, C.~H., Lejeune, T., \& Barnes, S.\ 2001, \apjs, 136, 417 
\bibitem[Zaritsky et al.(1994)]{Zaritsky1994} Zaritsky, D., 
Kennicutt, R.~C., Jr., \& Huchra, J.~P.\ 1994, \apj, 420, 87 


\end{thebibliography}
\end{document}